\documentclass[12pt,letterpaper]{article}
\usepackage{epsfig,rotating,setspace,latexsym,amsmath,epsf,amssymb,amsfonts,bm,theorem,cite,caption,subcaption,enumerate,longtable,accents}
\usepackage{algorithm,algorithmic,graphicx,epsf,authblk,epstopdf,url,color,multirow}
\newtheorem{theorem}{Theorem}

\newtheorem{corollary}{Corollary}

\newtheorem{remark}{Remark}
\newtheorem{lemma}{Lemma}
\newenvironment{Proof}[1]{\medskip\par\noindent{\bf Proof:\,}\,#1}{{\mbox{\,$\blacksquare$}\par}}

\newcommand{\cs}{{\mathcal{S}}}
\newcommand{\bt}{{\boldsymbol{\tau}}}

\allowdisplaybreaks

\begin{document}
	
\title{Asymmetry Hurts: Private Information Retrieval Under Asymmetric Traffic Constraints\thanks{This work was supported by NSF Grants  CNS 13-14733, CCF 14-22111, CNS 15-26608 and CCF 17-13977. A shorter version is submitted to IEEE ISIT 2018.}}
	
\author{Karim Banawan \qquad Sennur Ulukus\\
	\normalsize Department of Electrical and Computer Engineering\\
	\normalsize University of Maryland, College Park, MD 20742 \\
	\normalsize {\it kbanawan@umd.edu} \qquad {\it ulukus@umd.edu}}
	
\maketitle
	
\vspace*{-0.5cm}

\begin{abstract}
We consider the classical setting of private information retrieval (PIR) of a single message (file) out of $M$ messages from $N$ distributed databases under the new constraint of \emph{asymmetric traffic} from databases. In this problem, the \emph{ratios between the traffic} from the databases are constrained, i.e., the ratio of the length of the answer string that the user (retriever) receives from the $n$th database to the total length of all answer strings from all databases is constrained to be $\tau_n$. This may happen if the user's access to the databases is restricted due database availability, channel quality to the databases, and other factors. For this problem, for fixed $M$, $N$, we develop a general upper bound $\bar{C}(\bt)$, which generalizes the converse proof of Sun-Jafar, where database symmetry was inherently used. Our converse bound is a piece-wise affine function in the traffic ratio vector $\bt=(\tau_1, \cdots, \tau_N)$. For the lower bound, we explicitly show the achievability of $\binom{M+N-1}{M}$ corner points. For the remaining traffic ratio vectors, we perform time-sharing between these corner points. The recursive structure of our achievability scheme is captured via a system of difference equations. The upper and lower bounds exactly match for $M=2$ and $M=3$ for any $N$ and any $\bt$. The results show strict loss of PIR capacity due to the asymmetric traffic constraints compared with the symmetric case of Sun-Jafar which implicitly uses $\tau_n=\frac{1}{N}$ for all $n$.
\end{abstract}

\section{Introduction}
Protecting the privacy of downloaded information from curious publicly accessible databases has been the focus of considerable research within the computer science community \cite{ChorPIR, PIRsurvey2004, cachin1999computationally, ostrovsky2007survey, yekhanin2010private}. The problem of privacy has become even more relevant today in the presence of efficient data-mining techniques. Private information retrieval (PIR), introduced by Chor et al. in \cite{ChorPIR}, studies the privacy of the downloaded content from public databases. In the classical PIR setting, a user requests to download a certain message (or file) out of $M$ distinct messages from $N$ non-communicating (non-colluding) databases without leaking the identity of the desired message to any individual database. The contents of these databases are identical. The user prepares $N$ queries, one for each database, such that the queries do not reveal the user's interest in the desired message. Upon receiving these queries, each database responds truthfully with an answering string. The user needs to be able to reconstruct the entire message by decoding the answer strings from all databases. PIR schemes are designed to be more efficient than the trivial scheme of downloading all the files stored in the databases. The efficiency of a retrieval scheme is measured by the retrieval rate, which is the ratio of the number of decodable desired message symbols to the number of total downloaded symbols.

Recently, the PIR problem is revisited by information theorists \cite{RamchandranPIR, unsynchonizedPIR, YamamotoPIR, VardyConf2015, RazanPIR, JafarConf2016}. The information-theoretic reformulation of the problem assumes that the messages are of arbitrarily large size and hence the upload cost can be neglected with respect to the download cost \cite{YamamotoPIR} in contrast to the computer science formulation. This formulation provides an absolute guarantee (as opposed to computational PIR, e.g., \cite{yekhanin2010private, cachin1999computationally}). In the leading work \cite{JafarPIR}, Sun and Jafar introduce the PIR capacity notion to characterize the fundamental limits of the PIR problem. The PIR capacity is defined as the supremum of PIR rates over all achievable retrieval schemes. \cite{JafarPIR} determines the exact capacity of the classical PIR to be $C=(1+\frac{1}{N}+\frac{1}{N^2}+\cdots+\frac{1}{N^{M-1}})^{-1}$. Following the work of \cite{JafarPIR}, the fundamental limits of many interesting variants of the classical PIR problem have been considered, such as: PIR from colluding databases, robust PIR, symmetric PIR, PIR from MDS-coded databases, PIR for arbitrary message lengths, multi-round PIR, multi-message PIR, PIR from Byzantine databases, secure symmetric PIR with adversaries, cache-aided PIR, PIR with private side information (PSI), PIR for functions, storage constrained PIR, and their several combinations
\cite{JafarColluding, symmetricPIR, KarimCoded, arbmsgPIR, codedsymmetric, MultiroundPIR, codedcolluded, codedcolludedJafar, arbitraryCollusion, MPIRjournal, codedcolludingZhang, MPIRcodedcolludingZhang, BPIRjournal, symmetricByzantine, tandon2017capacity, wang2017linear, kadhe2017private, wei2017fundamental, chen2017capacity, wei2017capacity, sun2017_computation, mirmohseni2017private, abdul2017private,wei2017fundamental_partial}.

A common property of the achievability schemes constructed for these PIR problems is that they exhibit a \emph{symmetric structure} across the databases. In most existing PIR schemes, the user retrieves pieces of the desired message from all databases, and generates and uses side information at all databases in a symmetric manner. This enables the user to balance the load of retrieval of the desired message equally among the databases, and re-use the side information generated from one database equally in all the remaining databases. Now, consider the following scenarios that render symmetry assumption unworkable: {\it Varying database availability:} Certain databases are available only a fraction of the time other databases are available for downloads. {\it Different capacities:} The capacities of the links (bit pipes) from the databases to the user have different capacities. This may be due to different physical locations of the databases, e.g., the user may be able to access physically closer databases more often than physically distant databases, or it may be due to the quality of the physical layer communication channel, e.g., the bandwidths (rates) of the download channels may be different for different databases. In these cases, the user is forced to deal with each database differently, i.e., the user should utilize the databases which have better quality links more often than the other databases. This breaks the database symmetry assumption, makes load balancing of desired message and side information more challenging, and poses the following interesting questions: Can we perform efficient PIR without applying database symmetry? Is there a fundamental PIR rate loss due to not being able to use symmetric schemes?

Motivated by these practical scenarios, we consider the PIR problem under \emph{asymmetric traffic constraints}. Formally, we consider a classical PIR setting with $N$ replicated and non-communicating databases storing $M$ messages. We assume that the $n$th database responds with a $t_n$-length answer string. We constrain the lengths of the answer strings such that $t_n=\lambda_n t_1$ for $n \in \{2, \cdots, N\}$. This, in turn, forces the ratios between the traffic from the databases to be $1:\lambda_2:\lambda_3: \cdots: \lambda_N$. We denote the traffic ratio with respect to the total download by a vector $\bt=(\tau_1, \cdots, \tau_N)$, where $\tau_n=\frac{\lambda_n}{\sum_{j=1}^{N} \lambda_j}$. We aim at characterizing the capacity of this PIR problem, $C(\bt)$, as a function of the given traffic ratio vector $\bt$ for arbitrary $M$ and $N$. We note that in this problem, we do not constrain $t_1$ itself, but rather constrain the ratios between the responses according to $\bt$; in fact, we assume that $t_1$ can grow arbitrarily large to conform with the classical information-theoretic formulation. Furthermore, we remark that although our problem seems to be related to the \emph{upload-constrained} PIR problem \cite{JafarPIR}, we note that the upload-constrained problem investigates the \emph{minimum} possible query size if the user and the databases exchange a codebook prior to the retrieval process, while in the asymmetric traffic constrained problem here we do not assume the existence of a codebook, and hence we minimize the number of queries subject to an additional constraint on the traffic ratios.

In this paper, we investigate the fundamental limits of the PIR problem under asymmetric traffic constraints. To that end, we develop a novel upper bound for the capacity $\bar{C}(\bt)$. This generalizes the converse proof of \cite{JafarPIR} to incorporate the asymmetric traffic constraints. Originally, the proof in \cite{JafarPIR} exploits the database symmetry. The rationale is that even if the optimal scheme is not symmetric, we can transform it into a symmetric scheme without changing the retrieval rate by means of time-sharing \cite{JafarPIR}. In our case, we cannot use this technique as we must deal with the databases differently. We characterize the upper bound as a piece-wise affine function in $\bt$ (see Theorem~\ref{Thm1}). The upper bound implies that asymmetry fundamentally hurts the retrieval rate (see Corollary~\ref{cor1} and Remark~\ref{rem4}). Then, we propose explicit achievability schemes for $\binom{M+N-1}{M}$ corner points. Each corner point corresponds to a specific partitioning of the databases according to the number of side information symbols that are used simultaneously within the initial round of the download. We describe the achievability scheme via a system of difference equations in the number of stages at each round of the download (which is parallel to \cite{MPIRjournal}). For any other traffic ratio vector $\bt$, we employ time-sharing between the corner points that enclose $\bt$. We provide an explicit rate expression for the case of $N=2$ for arbitrary $M$. We show that the upper bound and the lower bound exactly match for the cases of $M=2$ and $M=3$ messages for any $N$ and any $\bt$, leading to the exact capacity $C(\bt)$ for these cases.

\section{System Model}
Consider a classical PIR model with $N$ non-communicating and replicated databases storing $M$ messages (or files). Each database stores the same set of messages $W_{1:M}=\{W_1, \cdots, W_M\}$. Messages $W_{1:M}$ are independent and identically distributed over all vectors of size $L$ picked from a finite field $\mathbb{F}_q^L$, i.e.,
\begin{align}
H(W_i)&=L, \quad i \in \{1, \cdots, M\} \\
H(W_1, \cdots, W_M)&=ML, \quad (q\text{-ary units})
\end{align}

In the PIR problem, a user wants to retrieve a message $W_i \in W_{1:M}$ correctly without revealing any information about the identity of the message $i$ to any individual database. To that end, the user submits a query $Q_n^{[i]}$ to the $n$th database. The messages and the queries are statistically independent due to the fact that the user does not know the message realizations in advance, i.e.,
\begin{align}\label{independency}
I(W_{1:M};Q_{1:N}^{[i]})=0
\end{align}
where $Q_{1:N}^{[i]}=\{Q_1^{[i]}, \cdots, Q_N^{[i]}\}$. The $n$th database responds truthfully by an answer string $A_n^{[i]}$. The answer string $A_n^{[i]}$ is a deterministic function of the query $Q_n^{[i]}$ and all the messages $W_{1:M}$, hence
\begin{align}\label{answer_constraint}
H(A_n^{[i]}|Q_n^{[i]}, W_{1:M})=0, \quad n \in \{1, \cdots, N\}
\end{align}

\begin{figure}[t]
	\centering
	\includegraphics[width=0.75\textwidth]{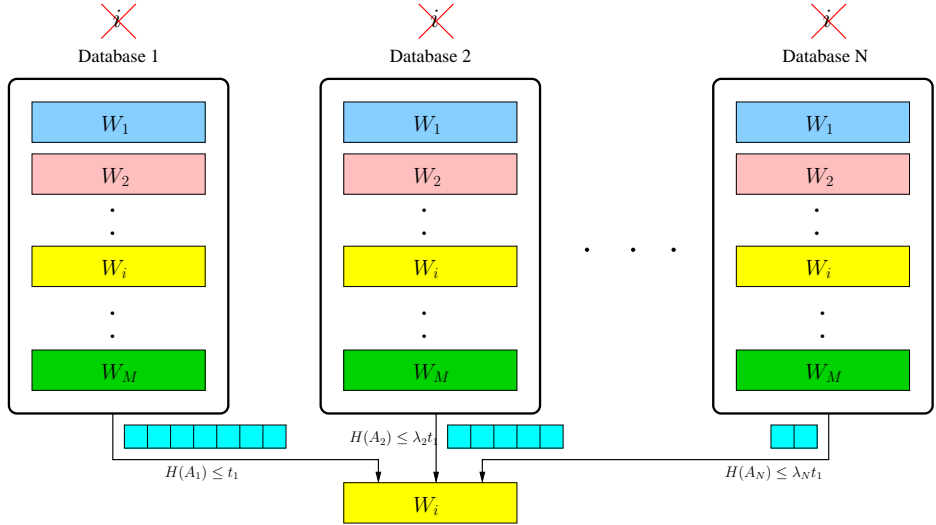}
	\caption{PIR under asymmetric traffic constraints.}
	\label{Fig:asymmetric-traffic}
	\vspace*{-0.4cm}
\end{figure}

In the PIR model with asymmetric traffic constraints, the lengths of the answer strings are different (see Fig.~\ref{Fig:asymmetric-traffic}). More specifically, we assume that the $n$th database responds with a $t_n$-length answer string, such that $t_n=\lambda_n t_1$, where $\lambda_n$ is the ratio between the traffic from the $n$th database to the traffic from the first database. Without loss of generality, we assume that the first database has the highest traffic and the remaining databases are ordered descendingly in $\lambda_n$. Hence, $\{\lambda_n\}_{n=1}^N$ is a \emph{non-increasing monotone} sequence with $\lambda_1=1$, and $\lambda_n \in [0,1]$, i.e.,
\begin{align}\label{asymmetric_traffic}
H(A_n^{[i]})&\leq \lambda_n t_1, \quad i \in \{1, \cdots, M\}, \: n \in \{1, \cdots, N\}, \: 1 \geq \lambda_2 \geq \cdots \geq \lambda_N
\end{align}

We define the \emph{traffic ratio} of the $n$th database $\tau_n$ as the ratio between the traffic from the $n$th database and the total traffic from all databases, i.e.,
\begin{align}
\tau_n=\frac{\lambda_n}{\sum_{j=1}^{N} \lambda_j}
\end{align}
We note that there is a one-to-one transformation between the vector $\boldsymbol{\lambda}=(\lambda_1,\lambda_2, \cdots, \lambda_N)$ and the vector $\boldsymbol{\tau}=(\tau_1, \tau_2, \cdots, \tau_N)$. Thus, $\boldsymbol{\lambda}$ and $\bt$ are used interchangeably within the context of this paper.

In order to ensure the privacy, at the $n$th database, the query $Q_n^{[i]}$ designed to retrieve $W_i$ should be indistinguishable from the queries designed to retrieve any other message, i.e.,
\begin{align}\label{privacy_constraint}
(Q_n^{[i]}, A_n^{[i]}, W_{1:M}) \sim (Q_n^{[j]}, A_n^{[j]}, W_{1:M}), \quad \forall j \in \{1, \cdots, M\}
\end{align}
where $\sim$ denotes statistical equivalence.

In addition, the user should be able to reconstruct $W_i$ from the collected answer strings $A_{1:N}^{[i]}$ with arbitrarily small probability of error. By Fano's inequality, we have the following reliability constraint,
\begin{align}\label{reliability_constraint}
H(W_i|Q_{1:N}^{[i]},A_{1:N}^{[i]})=o(L)
\end{align}
where $\frac{o(L)}{L} \rightarrow 0$ as $L \rightarrow \infty$.

For a fixed $N$, $M$, and a traffic ratio vector $\boldsymbol{\tau}$, a retrieval rate $R(\boldsymbol{\tau})$ is achievable if there exists a PIR scheme which satisfies the privacy constraint \eqref{privacy_constraint} and the reliability constraint \eqref{reliability_constraint} for some message lengths $L(\boldsymbol{\tau})$ and answer strings of lengths $\{t_n(\boldsymbol{\tau})\}_{n=1}^N$ that satisfy the asymmetric traffic constraint \eqref{asymmetric_traffic}, such that
\begin{align}
R(\boldsymbol{\tau})=\frac{L(\boldsymbol{\tau})}{\sum_{n=1}^N t_n(\boldsymbol{\tau})}
\end{align}

We note that in this problem, we do not constrain either the message length $L(\boldsymbol{\tau})$ or the lengths of the answer strings $t_n(\boldsymbol{\tau})$, but we rather constrain the ratios of the traffic of each database with respect to the traffic of the first database. The pair $(L(\boldsymbol{\tau}),t_1(\boldsymbol{\tau}))$ can grow arbitrarily large to conform with the information-theoretic framework.

The capacity of the PIR problem under asymmetric traffic constraints $C(\boldsymbol{\tau})$ is defined as the supremum of all achievable retrieval rates, i.e., $C(\boldsymbol{\tau})=\sup \:R(\boldsymbol{\tau})$.

\section{Main Results and Discussions}
Our first result is an upper bound on $C(\bt)$ as a function of $\bt$ for any fixed $M$, $N$.

\begin{theorem}[Upper bound]\label{Thm1}
	For the PIR problem under monotone non-increasing asymmetric traffic constraints $\bt=(\tau_1, \cdots, \tau_N)$, the PIR capacity  $C(\bt)$ is upper bounded by
	\begin{align}\label{upper_bound}
		C(\bt) \leq \bar{C}(\bt)= \min_{n_1, \cdots, n_{M-1} \in \{1, \cdots, N\}} \frac{1+\frac{\sum_{n=n_1+1}^{N} \tau_n}{n_1}+\frac{\sum_{n=n_2+1}^{N} \tau_n}{n_1n_2}+\cdots+\frac{\sum_{n=n_{M-1}+1}^{N} \tau_n}{n_0 n_1 \cdots n_{M-1}}}{1+\frac{1}{n_1}+\frac{1}{n_1n_2}+\cdots+\frac{1}{n_0 n_1 \cdots n_{M-1}}}
	\end{align}
\end{theorem}

The proof of this upper bound is given in Section~\ref{converse}. We have the following remarks.

\begin{remark}
	The minimization in (\ref{upper_bound}) is performed to obtain the tightest bound, i.e., the bound in \eqref{upper_bound} is valid for any sequence of $\{n_i\}_{i=1}^{N} \subset \{1, \cdots, N\}^{M-1}$. In particular, restricting the minimization in the bound in \eqref{upper_bound} to monotone non-decreasing sequences $\{n_i\}_{i=1}^{M-1} \subset \{1, \cdots, N\}^{M-1}$ such that $n_1 \leq n_2 \leq \cdots \leq n_{M-1}$ is still a valid upper bound. For fixed $M$, $N$, the number of such monotone bounds is $\binom{M+N-2}{M-1}$.
\end{remark}

\begin{remark}
	The upper bound for the capacity function $\bar{C}(\bt)$ in (\ref{upper_bound}) is a piece-wise affine function in the traffic ratio vector $\bt$.
\end{remark}

\begin{remark}
	The upper bound in \eqref{upper_bound} generalizes the known results about the PIR problem.  By picking $n_1=\cdots=n_{M-1}=N$, \eqref{upper_bound} leads to
	\begin{align}
	C(\bt) \leq \frac{1}{1+\frac{1}{N}+\frac{1}{N^2}+\cdots+\frac{1}{N^{M-1}}}
	\end{align}
	which is the capacity of PIR with symmetric traffic (no traffic constraints) in \cite{JafarPIR}. On the other hand, if $\bt=(1, 0, 0, \cdots, 0)$, which implies that no traffic is returned from any database except for the first one, by picking $n_1= \cdots=n_{M-1}=1$, the upper bound in \eqref{upper_bound} leads to $\frac{1}{M}$, which is the capacity of the PIR problem with one database \cite{ChorPIR}.
\end{remark}

The following corollary is a direct consequence of Theorem~\ref{Thm1}. The corollary asserts that there is a strict capacity loss due to the asymmetric traffic constraints if the traffic ratio of the weakest link falls below a certain threshold.

\begin{corollary}[Asymmetry hurts]\label{cor1}
	For the PIR problem under  monotone non-increasing asymmetric traffic constraints $\bt=(\tau_1, \cdots, \tau_N)$, if $\tau_N<\tau^*$, such that
	\begin{align}\label{cor1-eqn}
	\tau^*=\frac{N^{M-1}-1}{N^M-1}, \quad N>1
	\end{align}
	then $C(\bt) < C$, where $C=\frac{1}{1+\frac{1}{N}+\cdots+\frac{1}{N^{M-1}}}$ is the PIR capacity without the asymmetric traffic constraints in \cite{JafarPIR}.
\end{corollary}

\begin{Proof}
	From Theorem~\ref{Thm1}, the upper bound corresponding to $n_1=N-1$, and $n_2=\cdots=n_{M-1}=N$ is strictly tighter than the capacity without asymmetric traffic constraints $C$ if
	\begin{align} \frac{1+\frac{\tau_N}{N-1}}{1+\frac{1}{N-1}+\frac{1}{(N-1)N}+\cdots+\frac{1}{(N-1)N^{M-1}}} &< C
	\end{align}
	which leads to
	\begin{align}
      \frac{\tau_N}{N-1}\left(1+\frac{1}{N}+\cdots+\frac{1}{N^{M-1}}\right)&<\left(\frac{1}{N-1}-\frac{1}{N}\right)\left(1+\frac{1}{N}+\cdots+\frac{1}{N^{M-2}}\right)
	\end{align}
	which further simplifies to
	\begin{align}
	\tau_N < \frac{\frac{1}{N}\left(1+\frac{1}{N}+\cdots+\frac{1}{N^{M-2}}\right)}{\left(1+\frac{1}{N}+\cdots+\frac{1}{N^{M-1}}\right)}=\frac{\sum_{i=0}^{M-2} N^i}{\sum_{i=0}^{M-1} N^i}=\tau^*
	\end{align}
	which implies that the upper bound for the capacity under the asymmetric traffic constraint is strictly less than $C$, which in turn implies that any achievable rate is strictly less than the unconstrained capacity.
\end{Proof}

\begin{remark}\label{rem4}
	As the number of messages $M$ becomes large enough, i.e., as $M \rightarrow \infty$, the traffic ratio threshold in (\ref{cor1-eqn}) $\tau^* \rightarrow \frac{1}{N}$. This implies that as $M \rightarrow \infty$, any asymmetric traffic constraint incurs strict capacity loss.
\end{remark}

Our second result is a lower bound on $C(\bt)$ as a function of $\bt$ for any fixed $M$, $N$.

\begin{theorem}[Lower bound]\label{Thm2}
	For the PIR problem under asymmetric traffic constraints, for a monotone non-decreasing sequence $\mathbf{n}=\{n_i\}_{i=0}^{M-1} \subset \{1, \cdots, N\}^{M}$, let $n_{-1}=0$, and $\cs=\{i \geq 0: n_i-n_{i-1}>0\}$. Denote $y_\ell[k]$ as the number of stages of the achievable scheme that downloads $k$-sums from the $n$th database, such that $n_{\ell-1} \leq n \leq n_{\ell}$, and $\ell \in \cs$. Let $\xi_\ell=\prod_{s \in \cs \setminus \{\ell\}} \binom{M-2}{s-1}$. The number of stages $y_\ell[k]$ is characterized by the following system of difference equations:
	\begin{align}\label{difference_eqn}
	y_0[k]&=(n_0\!-\!1)y_0[k\!-\!1]+\sum_{j \in \cs \setminus \{0\}} (n_j\!-\!n_{j-1}) y_j[k\!-\!1] \notag\\
	y_1[k]&=(n_1\!-\!n_0\!-\!1)y_1[k\!-\!1]+\sum_{j \in \cs \setminus \{1\}} (n_j\!-\!n_{j-1}) y_j[k\!-\!1] \notag\\
	y_\ell[k]&=n_0 \xi_\ell \delta[k\!-\!\ell\!-\!1]+(n_\ell\!-\!n_{\ell-1}\!-\!1) y_\ell[k-1]+\sum_{j \in \cs \setminus \{\ell\}} (n_j\!-\!n_{j-1})y_j[k\!-\!1], \quad  \ell \geq 2
	\end{align}
	where $\delta[\cdot]$ denotes the Kronecker delta function. The initial conditions of \eqref{difference_eqn} are $y_0[1]=\prod_{s \in \cs} \binom{M-2}{s-1}$, and $y_j[k]=0$ for $k \leq j$. Consequently, the traffic ratio vector $\bt(\mathbf{n})=(\tau_1(\mathbf{n}), \cdots, \tau_N(\mathbf{n}))$ corresponding to the sequence $\mathbf{n}=\{n_i\}_{i=0}^{M-1}$ is given by:
	\begin{align}
	\tau_n(\mathbf{n})=\frac{\sum_{k=1}^M \binom{M}{k} y_j[k]}{\sum_{\ell \in \cs} \sum_{k=1}^M \binom{M}{k} y_\ell[k] (n_\ell-n_{\ell-1})},\quad n_{j-1}+1 \leq n \leq n_j
	\end{align}
	and the achievable rate corresponding to $\bt(\mathbf{n})$ is given by:
	\begin{align}
	R(\bt(\mathbf{n}))=\frac{\sum_{\ell \in \cs} \sum_{k=1}^M \binom{M-1}{k-1} y_\ell[k] (n_\ell-n_{\ell-1})}{\sum_{\ell \in \cs} \sum_{k=1}^M \binom{M}{k} y_\ell[k] (n_\ell-n_{\ell-1})}
	\end{align}
	Moreover, for $\bt=\sum_{i=1}^{N} \alpha_i \bt\left(\mathbf{n}_i\right)$ for $\alpha_i \geq 0$, for all $i$, and $\sum_{i=1}^{N} \alpha_i=1$, the following is a lower bound on $C(\bt)$,
	\begin{align}
	C(\bt) \geq R(\bt) = \sum_{i=1}^{N} \alpha_i R(\bt(\mathbf{n}_i))
	\end{align}
\end{theorem}

The proof of Theorem~\ref{Thm2} can be found in Section~\ref{achievability}. The theorem characterizes an achievable rate for the corner points $\bt(\mathbf{n})$ corresponding to any monotone non-decreasing sequence $\mathbf{n}=\{n_i\}_{i=0}^{M-1} \subset \{1, \cdots, N\}^{M}$. For any other traffic ratio vector $\bt$, the achievability scheme is obtained by time-sharing between the nearest corner points. We note that due to the large number of corner points, we do not provide an explicit achievable rate for each corner point but we rather describe the achievable rate by a system of difference equations. The solution of this system of difference equations specifies the traffic ratio vector $\bt(\mathbf{n})$ and the achievable rate $R(\bt(\mathbf{n}))$ corresponding to the monotone non-decreasing sequence $\{n_i\}_{i=0}^{M-1}$. We have the following remarks.

\begin{remark}
	If $n_i=N$ for all $i \in \{0, \cdots, M-1\}$, then $\cs=\{0\}$ and the number of stages of $k$-sums is described by the following difference equation for any database
	\begin{align}
	y[k]=(N-1)y[k-1]
	\end{align}
	with initial condition of $y[1]=1$. In this case $\tau_n=\frac{1}{N}$ for all $n$, and $R=\frac{1}{1+\frac{1}{N}+\cdots+\frac{1}{N^{M-1}}}$, i.e., the scheme in Theorem~\ref{Thm2} reduces to the symmetric scheme in \cite{JafarPIR} if the sequence $\mathbf{n}=(N,N, \cdots, N)$ is used.
\end{remark}

\begin{remark}
     We note that the sequence $\{n_i\}_{i=0}^{M-1}$ suffices to completely specify the traffic ratio vector $\bt(\mathbf{n})$ for every corner point as a consequence of the monotonicity of the sequence, i.e.,
     \begin{align}
     \{n_i\}_{i=0}^{M-1} \Rightarrow (\underbrace{\tilde{\tau}_0, \cdots, \tilde{\tau}_0}_{n_0\: \text{elements}}, \underbrace{\tilde{\tau}_1, \cdots, \tilde{\tau}_1}_{(n_1-n_0)\: \text{elements}}, \cdots, \underbrace{\tilde{\tau}_{M-1}, \cdots, \tilde{\tau}_{M-1}}_{(n_{M-1}-n_{M-2})\: \text{elements}})
     \end{align}
     where $\tilde{\tau}_j=\frac{\sum_{k=1}^M \binom{M}{k} y_j[k]}{\sum_{\ell \in \cs} \sum_{k=1}^M \binom{M}{k} y_j[k] (n_\ell-n_{\ell-1})}$.
\end{remark}

\begin{remark}
For fixed $M$, $N$, the number of corner points in Theorem~\ref{Thm2} corresponds to the number of monotone non-decreasing sequences $\mathbf{n}=\{n_i\}_{i=0}^{M-1}$, which is $\binom{M+N-1}{M}$.
\end{remark}

The next corollary asserts that the achievable scheme in Theorem~\ref{Thm2} is optimal for $M=2$ and $M=3$ messages for any traffic ratio vector $\bt$ and any number of databases $N$.

\begin{corollary}[Capacity for $M=2$ and $M=3$ messages]\label{optimality23}
	For the PIR problem with asymmetric traffic constraints $\bt$, the capacity $C(\bt)$ for $M=2$ and $M=3$, and for any arbitrary $N$ is given by:
	\begin{align}\label{capacityM32}
	C(\bt) =
	\left\{
	\begin{array}{ll}
	\min_{n_0 \in \{1, \cdots, N\}} \frac{1+\frac{1}{n_0}\sum_{n=n_0+1}^{N} \tau_n}{1+\frac{1}{n_0}}, & M=2 \\
	\min_{n_0 \leq n_1 \in \{1, \cdots, N\}} \frac{1+\frac{1}{n_0}\sum_{n=n_0+1}^{N} \tau_n+\frac{1}{n_0n_1}\sum_{n=n_1+1}^{N} \tau_n}{1+\frac{1}{n_0}+\frac{1}{n_0n_1}}, & M=3
	\end{array}
	\right.
	\end{align}
\end{corollary}

The proof of Corollary~\ref{optimality23} is given in Section~\ref{optimal_achievability}.

\begin{figure}[t]
	\centering
	\includegraphics[width=0.75\textwidth]{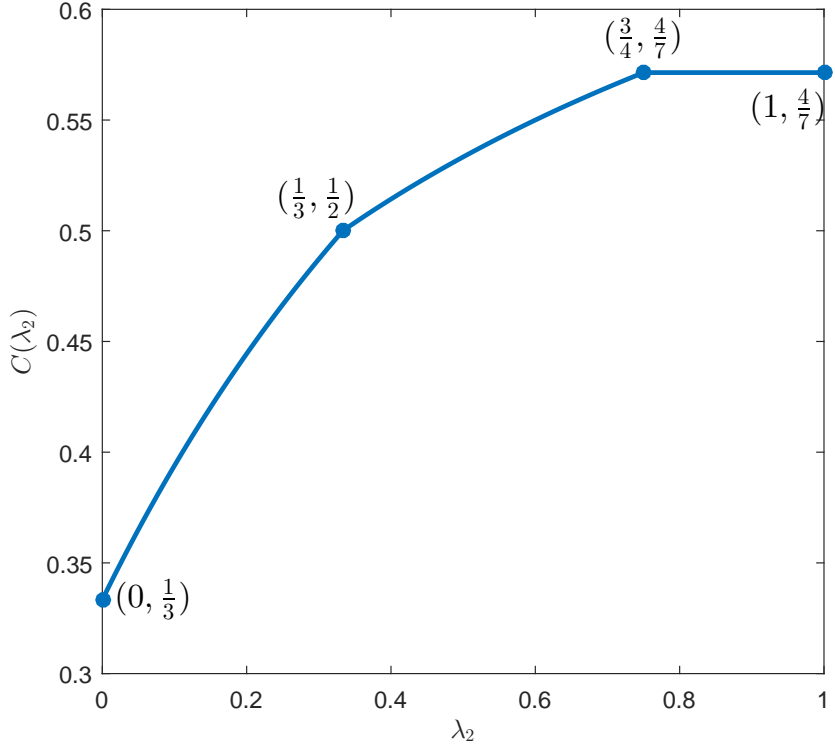}
	\caption{Capacity function $C(\lambda_2)$ for $M=3$, $N=2$.}
	\label{Fig:M3N2}
	\vspace*{-0.4cm}
\end{figure}

Fig.~\ref{Fig:M3N2} shows the PIR capacity under asymmetric constraints $C(\lambda_2)$ as a function of $\lambda_2$ (which is bijective to $\bt$) for the case of $M=3$ messages and $N=2$ databases. We note that the capacity $C(\lambda_2)$ is a piece-wise monotone curve in $\lambda_2$, which consists of $\binom{M+N-2}{M-1}=3$ regimes. There exist $\binom{M+N-1}{M}=4$ corner points. Specific achievable schemes for the case of $M=3$ and $N=2$ are given in Section~\ref{motivating-ex}. Each corner point shown in Fig.~\ref{Fig:M3N2} corresponds to an explicit achievable scheme given in Section~\ref{M3N2 corner}. For any other point, time-sharing between corner points is used to achieve these points as shown in Section~\ref{M3N2 non-corner}.

\begin{figure}[t]
	\centering
	\includegraphics[width=0.9\textwidth]{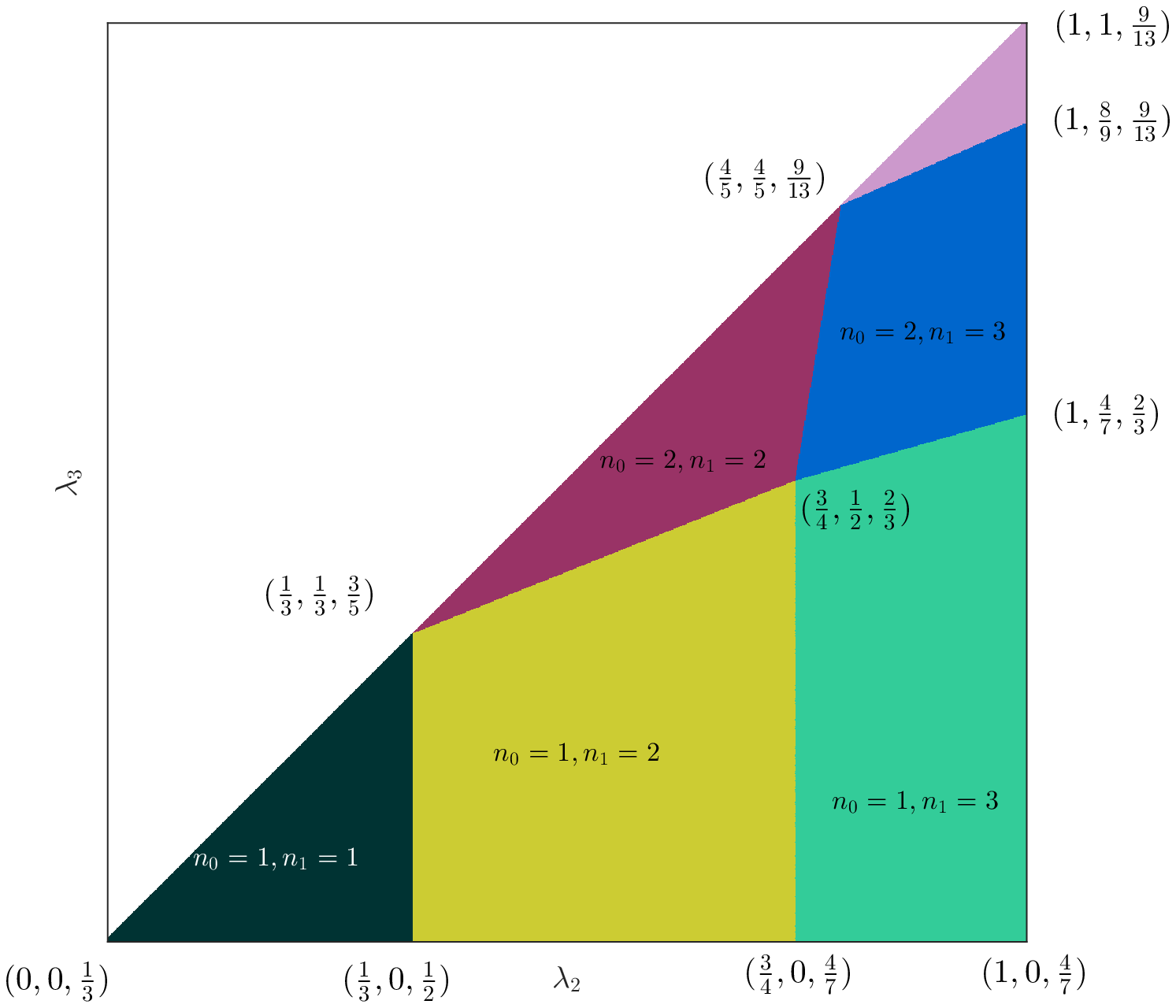}
	\caption{Illustration of corner points and regions of $C(\lambda_2,\lambda_3)$ for $M=3$, $N=3$.}
	\label{Fig:M3N3}
	\vspace*{-0.4cm}
\end{figure}

Fig.~\ref{Fig:M3N3} shows the capacity region $C(\lambda_2,\lambda_3)$ for the case of $M=3$ messages and $N=3$ databases as a function of the pair $(\lambda_2,\lambda_3)$ (which is bijective to $\bt$). Fig.~\ref{Fig:M3N3} shows that there exist $\binom{M+N-1}{M}=10$ corner points, and $\binom{M+N-2}{M-1}=6$ regions. We show the capacity regions in terms of the triple $(\lambda_2,\lambda_3, C(\lambda_2,\lambda_3))$. Furthermore, for every region we show the corresponding $(n_0,n_1)$ to be plugged in \eqref{capacityM32}. The capacity for any point $(\lambda_2,\lambda_3)$ other than the corner points is obtained by time-sharing between the corner points that enclose $(\lambda_2,\lambda_3)$. Specific achievable schemes for $M=3$, $N=3$ are given in Section~\ref{M3N3_scheme}.

Finally, in the following corollary, we specialize the achievable scheme in Theorem~\ref{Thm2} to the case of $N=2$ for any arbitrary $M$.

\begin{corollary}[Achievable traffic versus retrieval rate tradeoff for $N=2$ databases]\label{corollary:N=2}
	For the PIR problem with $N=2$ and an arbitrary $M$ under asymmetric traffic constraints $\bt=(1-\tau_2,\tau_2)$, $\tau_2 \leq \frac{1}{2}$, let $s_2 \in \{1, \cdots, M-1\}$, for the traffic ratio $\tau_2(s_2)$, where
	\begin{align}
	\tau_2(s_2)=\frac{\sum_{i=0}^{\left\lfloor\frac{M-s_2-1}{2}\right\rfloor} \binom{M}{s_2+2i+1}}{M\binom{M-2}{s_2-1}+\sum_{i=0}^{M-s_2-1} \binom{M}{s_2+1+i}}
	\end{align}
	the PIR capacity $C(\tau_2(s_2))$ is lower bounded by:
	\begin{align}
	C(\tau_2(s_2)) \geq R(\tau_2(s_2))=\frac{\binom{M-2}{s_2-1}+\sum_{i=0}^{M-s_2-1} \binom{M-1}{s_2+i}}{M\binom{M-2}{s_2-1}+\sum_{i=0}^{M-s_2-1} \binom{M}{s_2+1+i}}
	\end{align}
	Moreover, if $\tau_2(s_2) < \tau_2 < \tau_2(s_2+1)$, and $\alpha \in (0,1)$, such that $\tau_2=\alpha\tau_2(s_2)+(1-\alpha)\tau_2(s_2+1)$, then
	\begin{align}
	C(\tau_2) \geq R(\tau_2)=\alpha R(\tau_2(s_2))+(1-\alpha)R(\tau_2(s_2+1))
	\end{align}
\end{corollary}

The proof of Corollary~\ref{corollary:N=2} is given in Section~\ref{Tradeoff N=2}.

\begin{figure}[t]
	\centering
	\includegraphics[width=0.75\textwidth]{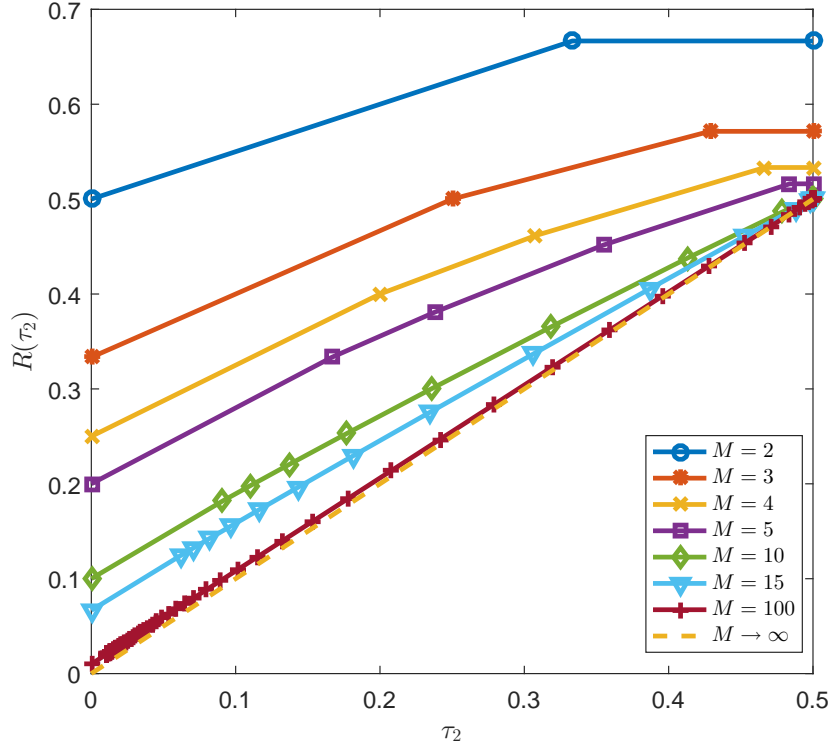}
	\caption{Achievable rate-traffic ratio tradeoff for $N=2$.}
	\label{Fig:N2}
	\vspace*{-0.4cm}
\end{figure}

\begin{remark}
Fig.~\ref{Fig:N2} shows the tradeoff between the traffic ratio $\tau_2$ and the achievable retrieval rate $R(\tau_2)$. We note that as $M$ increases $R(\tau_2)$ decreases pointwise. We observe that as $M \rightarrow \infty$, the rate-traffic tradeoff converges to $R(\tau_2)=\tau_2$. This implies that for large enough $M$, our achievable scheme reduces to time-sharing between the trivial achievable scheme of downloading all the messages from database 1 which achieves a rate of $\frac{1}{M}$, and the asymptotically-optimal achievable scheme in \cite{RamchandranPIR} which achieves $R=1-\frac{1}{N}$.
\end{remark}

\section{Converse Proof}\label{converse}
In this section, we derive an upper bound for the PIR problem with asymmetric traffic constraints. We extend the converse techniques introduced in \cite{JafarPIR} to account for the asymmetry of the returned answer strings.

We first need the following lemma, which characterizes a fundamental lower bound on the interference from the undesired messages within the answer strings, i.e., a lower bound on $\sum_{n=1}^{N} t_n-L$, as a consequence of the privacy constraint. The proof of this lemma can be found in \cite[Lemma~5]{JafarPIR}. The proof follows for our case since the privacy constraint does not change in the PIR with asymmetric traffic constraints, and the fact that the proof in \cite[Lemma~5]{JafarPIR} deals with the length of the entire downloaded answer strings $A_{1:N}^{[1]}$ and not the individual lengths of each answer string, see \cite[equations (46)-(47)]{JafarPIR}.
\begin{lemma}[Interference lower bound]\label{lemma_converse1}
	For the PIR problem under asymmetric traffic constraints $\{t_n\}_{n=1}^N$, the interference from undesired messages within the answer strings $\sum_{n=1}^{N} t_n-L$ is lower bounded as, 	
	\begin{align}
	 \sum_{n=1}^{N} t_n-L+ o(L) \geq I\left(W_{2:M};Q_{1:N}^{[1]}, A_{1:N}^{[1]}|W_{1} \right) \label{eq_L1}
	\end{align}
\end{lemma}

In the following lemma, we prove an inductive relation for the mutual information term on the right hand side of \eqref{eq_L1}. In this lemma, the interference lower bound in \eqref{eq_L1} is expanded into two parts. The first part, which contains the answer strings from the first $n_{m-1}$ databases $A_{1:n_{m-1}}^{[m]}$, is dealt with as in the proof of \cite[Lemma~6]{JafarPIR}. For the second part, which contains the remaining answer strings $A_{n_{m-1}+1:N}^{[m]}$, each answer string $A_n^{[m]}$ is bounded trivially by the length of the answer string $t_n$.
\begin{lemma}[Induction lemma]\label{lemma_converse2}
	For all $m\in \{2,\dots,M\}$ and for an arbitrary $n_{m-1} \in \{1, \cdots, N\}$, the mutual information term in Lemma~\ref{lemma_converse1} can be inductively lower bounded as,
	\begin{align} \label{eq_L2}
	&I\left( W_{m:M} ; Q_{1:N}^{[m-1]}, A_{1:N}^{[m-1]} | W_{1:m-1} \right)  \notag \\
	&\quad\quad\quad \geq \frac{1}{n_{m-1}}  I\left(W_{m+1:M}; Q_{1:N}^{[m]}, A_{1:N}^{[m]}|W_{1:m} \right)  +\frac{1}{n_{m-1}} \left(L-t_1\sum_{n=n_{m-1}+1}^N \lambda_n \right)-\frac{o(L)}{n_{m-1}}
	\end{align}
\end{lemma}

We note that \cite[Lemma~6]{JafarPIR} can be interpreted as a special case of Lemma~\ref{lemma_converse2} with setting $n_{m-1}=N$. Intuitively, $n_{m-1}$ represents the number of databases that can apply the optimal symmetric scheme in \cite{JafarPIR} if the user wants to retrieve message $W_{m-1}$ from the set of $W_{m-1:M}$ messages (i.e., conditioned on $W_{1:m-1}$).

\begin{Proof}
	We start with the left hand side of \eqref{eq_L2} after multiplying by $n_{m-1}$,
	\begin{align}
	& n_{m-1}\,I\left(W_{m:M} ; Q_{1:N}^{[m-1]}, A_{1:N}^{[m-1]}| W_{1:m-1}  \right) \notag \\
	&\qquad \geq  n_{m-1}\, I\left(W_{m:M} ; Q_{1:n_{m-1}}^{[m-1]}, A_{1:n_{m-1}}^{[m-1]}| W_{1:m-1}  \right) \label{eq_IL_12}\\
	& \label{eq_IL_1}\qquad \geq  \sum_{n=1}^{n_{m-1}} I\left(W_{m:M} ; Q_n^{[m-1]}, A_n^{[m-1]}| W_{1:m-1}  \right) \\
	& \label{eq_IL_2}\qquad \stackrel{\eqref{privacy_constraint}}{=}  \sum_{n=1}^{n_{m-1}} I\left(W_{m:M} ; Q_n^{[m]}, A_n^{[m]}| W_{1:m-1}  \right) \\
	& \label{eq_IL_444}\qquad \stackrel{\eqref{independency}}{=}  \sum_{n=1}^{n_{m-1}} I\left(W_{m:M} ;  A_n^{[m]}|Q_n^{[m]}, W_{1:m-1}  \right)\\
	& \label{eq_IL_44}\qquad \stackrel{\eqref{answer_constraint}}{=}  \sum_{n=1}^{n_{m-1}} H\left(A_n^{[m]}|Q_n^{[m]}, W_{1:m-1}  \right) \\
	& \label{eq_IL_3}\qquad \geq \sum_{n=1}^{n_{m-1}} H\left(A_n^{[m]}|A^{[m]}_{1:n-1},Q_{1:n_{m-1}}^{[m]}, W_{1:m-1}  \right) \\
	& \label{eq_IL_4}\qquad \stackrel{\eqref{answer_constraint}}{=} \sum_{n=1}^{n_{m-1}} I\left(W_{m:M};A_n^{[m]}|A^{[m]}_{1:n-1},Q_{1:n_{m-1}}^{[m]}, W_{1:m-1}  \right) \\
	&\label{eq_IL_5}  \qquad = I\left(W_{m:M}; A_{1:n_{m-1}}^{[m]} | Q_{1:n_{m-1}}^{[m]}, W_{1:m-1}  \right) \\
	&\label{eq_IL_6} \qquad \stackrel{\eqref{independency}}{=} I\left(W_{m:M}; Q_{1:n_{m-1}}^{[m]}, A_{1:n_{m-1}}^{[m]} |  W_{1:m-1}  \right)\\
	&\label{eq_IL_65}\quad\:\: \stackrel{\eqref{independency},\eqref{answer_constraint}}{=}I\left(W_{m:M}; Q_{1:N}^{[m]}, A_{1:N}^{[m]} |  W_{1:m-1}\right)\!-\!I\left(W_{m:M}; A_{n_{m-1}+1:N}^{[m]} | Q_{1:N}^{[m]}, A_{1:n_{m-1}}^{[m]}, W_{1:m-1}  \right)\\
	&\qquad \label{eq_IL_66}\stackrel{\eqref{answer_constraint}}{=} I\left(W_{m:M}; Q_{1:N}^{[m]}, A_{1:N}^{[m]} |  W_{1:m-1}\right)-H\left(A_{n_{m-1}+1:N}^{[m]} | Q_{1:N}^{[m]}, A_{1:n_{m-1}}^{[m]}, W_{1:m-1}  \right)\\
	&\qquad \label{eq_IL_67}\stackrel{\eqref{asymmetric_traffic}}{\geq}I\left(W_{m:M}; Q_{1:N}^{[m]}, A_{1:N}^{[m]} |  W_{1:m-1}\right)-t_1 \sum_{n=n_{m-1}+1}^{N} \lambda_n\\
	& \label{eq_IL_7} \qquad \stackrel{\eqref{reliability_constraint}}{=} I\left(W_{m:M}; W_m, Q_{1:N}^{[m]}, A_{1:N}^{[m]} |  W_{1:m-1}\right)-t_1 \sum_{n=n_{m-1}+1}^{N} \lambda_n-o(L)\\
	& \qquad =I\left(W_{m:M};W_m|  W_{1:m-1}\right)+I\left(W_{m:M};Q_{1:N}^{[m]}, A_{1:N}^{[m]} |  W_{1:m}\right)-t_1 \sum_{n=n_{m-1}+1}^{N} \lambda_n -o(L)\\
	& \qquad =I\left(W_{m+1:M}; Q_{1:N}^{[m]}, A_{1:N}^{[m]}|W_{1:m} \right)  + \left(L-t_1\sum_{n=n_{m-1}+1}^N \lambda_n \right)-o(L)
	\end{align}
where \eqref{eq_IL_12}, \eqref{eq_IL_1} follow from the non-negativity of mutual information, \eqref{eq_IL_2} follows from the privacy constraint, \eqref{eq_IL_444} follows from the independence of $\left(W_{m:M},Q_n^{[m]}\right)$, \eqref{eq_IL_44},\eqref{eq_IL_4} follow from the fact that the answer string $A_n^{[m]}$ is a deterministic function of $\left(Q_n^{[m]},W_{1:M}\right)$, \eqref{eq_IL_3} follows from conditioning reduces entropy, \eqref{eq_IL_6} follows from the independence of $\left(W_{m:M},Q_{1:n_{m-1}}^{[m]}\right)$, \eqref{eq_IL_65} follows from the chain rule, the independence of the queries and the messages, and the fact that $Q_{1:N}^{[m]} \rightarrow Q_{1:n_{m-1}}^{[m]} \rightarrow A_{1:n_{m-1}}^{[m]}$ forms a Markov chain by (\ref{answer_constraint}), \eqref{eq_IL_66} follows from the fact that the answer strings $A_{1:n_{m-1}}^{[m]}$ are fully determined from $\left(Q_{1:N}^{[m]},W_{1:M}\right)$, \eqref{eq_IL_67} follows from the fact that conditioning reduces entropy and $H(A_{n_{m-1}+1:N}) \leq \sum_{n=n_{m-1}+1}^N t_n$ which is equal to $t_1 \sum_{n=n_{m-1}+1}^N \lambda_n$ from the asymmetric traffic constraints, \eqref{eq_IL_7} follows from the reliability constraint. Finally, dividing both sides by $n_{m-1}$ leads to \eqref{eq_L2}.
\end{Proof}

Now, we are ready to derive an explicit upper bound for the retrieval rate under asymmetric traffic constraints. Applying Lemma~\ref{lemma_converse1} and Lemma~\ref{lemma_converse2} successively for an arbitrary sequence $\{n_i\}_{i=1}^{M-1} \subset \{1, \cdots, N\}^{M-1}$ and observing that $\sum_{n=1}^{N}t_n=t_1 \sum_{n=1}^{N} \lambda_n$ under the asymmetric traffic constraints, we have the following
\begin{align}
&t_1\sum_{n=1}^N \lambda_n-L+\tilde{o}(L)  \notag \\
&\label{eq_induction1}\quad \stackrel{\eqref{eq_L1}}{\geq} I\left(W_{2:M}; Q_{1:N}^{[1]}, A_{1:N}^{[1]}|W_{1} \right) \\
&\quad \stackrel{\eqref{eq_L2}}{\geq}\frac{1}{n_{1}} \left(\!L\!-\!t_1\!\sum_{n=n_{1}+1}^N \!\lambda_n \right)+ \frac{1}{n_{1}}  I\left(W_{3:M}; Q_{1:N}^{[2]}, A_{1:N}^{[2]}|W_{1:2} \right)    \\
&\quad \stackrel{\eqref{eq_L2}}{\geq}\frac{1}{n_{1}} \left(\!L\!-\!t_1\!\sum_{n=n_{1}+1}^N \!\lambda_n \right)+\frac{1}{n_{1}n_2} \left(\!L\!-\!t_1\!\sum_{n=n_{2}+1}^N \!\lambda_n \right)+ \frac{1}{n_{2}}  I\left(W_{4:M}; Q_{1:N}^{[3]}, A_{1:N}^{[3]}|W_{1:3} \right)    \\
&\quad \stackrel{\eqref{eq_L2}}{\geq} \dots \notag\\
&\quad \stackrel{\eqref{eq_L2}}{\geq}\frac{1}{n_{1}}\!\left(\!L\!-\!t_1\!\sum_{n=n_{1}+1}^N \!\lambda_n \right)\! +\!\frac{1}{n_{1}n_2}\!\left(\!L\!-\!t_1\!\sum_{n=n_{2}+1}^N \!\lambda_n \right)\!+\!  \cdots \!+\!\frac{1}{\prod_{i=1}^{M-1} n_i}\!\left(\!L\!-\!t_1\!\sum_{n=n_{M-1}+1}^N \!\!\!\!\!\lambda_n \right)
\end{align}
where $\tilde{o}(L)=\left(1+\frac{1}{n_1}+\frac{1}{n_1n_2}+\cdots+\frac{1}{\prod_{i=1}^{M-1}n_i}\right)o(L)$, \eqref{eq_induction1} follows from Lemma~\ref{lemma_converse1}, and the remaining bounding steps follow from successive application of Lemma~\ref{lemma_converse2}.

Ordering terms, we have,
\begin{align}
\left(1+\frac{1}{n_1}\!+\!\frac{1}{n_1n_2}\!+\!\cdots\!+\!\frac{1}{\prod_{i=1}^{M-1}n_i}\right)\!L \leq
\left(1+\frac{\gamma(n_1)}{n_1}+\!\cdots\!+\frac{\gamma(n_{M-1})}{\prod_{i=1}^{M-1} n_i}\right)t_1\!\sum_{n=1}^N \lambda_n \!+\! \tilde{o}(L)
\end{align}
where $\gamma(\ell)=\frac{\sum_{n=\ell+1}^{N}\lambda_n}{\sum_{n=1}^{N} \lambda_n}=\sum_{n=\ell+1}^{N} \tau_n$ corresponds to the sum of the traffic ratios from databases $[\ell+1:N]$.

We conclude the proof by taking $L \rightarrow \infty$. Thus, for an arbitrary sequence $\{n_i\}_{i=1}^{M-1}$, we have
\begin{align}
R(\boldsymbol{\tau})&= \frac{L}{t_1 \sum_{n=1}^{N} \lambda_n } \leq  \frac{1+\frac{\gamma(n_1)}{n_1}+\frac{\gamma(n_2)}{n_1n_2}+\cdots+\frac{\gamma(n_{M-1})}{\prod_{i=1}^{M-1} n_i}}{1+\frac{1}{n_1}+\frac{1}{n_1n_2}+\cdots+\frac{1}{\prod_{i=1}^{M-1}n_i}}
\end{align}
Finally, we get the tightest bound by minimizing over the sequence $\{n_i\}_{i=1}^{M-1}$ over the set $\{1, \cdots, N\}$, as

\begin{align}
R(\boldsymbol{\tau}) &\leq \min_{n_1, \cdots, n_{M-1} \in \{1, \cdots, N\}} \frac{1+\frac{\gamma(n_1)}{n_1}+\frac{\gamma(n_2)}{n_1n_2}+\cdots+\frac{\gamma(n_{M-1})}{\prod_{i=1}^{M-1} n_i}}{1+\frac{1}{n_1}+\frac{1}{n_1n_2}+\cdots+\frac{1}{\prod_{i=1}^{M-1}n_i}}\\
   &=\min_{n_1, \cdots, n_{M-1} \in \{1, \cdots, N\}} \frac{1+\frac{\sum_{n=n_1+1}^{N} \tau_n}{n_1}+\frac{\sum_{n=n_2+1}^{N} \tau_n}{n_1n_2}+\cdots+\frac{\sum_{n=n_{M-1}+1}^{N} \tau_n}{\prod_{i=1}^{M-1} n_i}}{1+\frac{1}{n_1}+\frac{1}{n_1n_2}+\cdots+\frac{1}{\prod_{i=1}^{M-1}n_i}}
\end{align}

\begin{remark}
	From the converse proof, we note that we can intuitively interpret $n_i$ as the number of databases that can apply the symmetric traffic scheme in \cite{JafarPIR} if the number of messages is reduced to be $M-i+1$. We point out that in the absence of asymmetric traffic constraints as in \cite{JafarPIR}, all databases can apply symmetric schemes, therefore $n_i=N$ for all $i \in \{1, \cdots, M-1\}$. Now, since reducing the number of messages cannot decrease the number of databases that apply the symmetric scheme as the problem would be less constrained (in terms of the privacy constraint), which leads to more flexibility in terms of satisfying the traffic constraints, it suffices to evaluate the bound in \eqref{upper_bound} for monotone non-decreasing sequences $\{n_i\}_{i=1}^{M-1} \subset \{1, \cdots, N\}^{M-1}$ such that $n_1 \leq n_2 \leq \cdots \leq n_{M-1}$.
\end{remark}

\section{Achievability Proof}\label{achievability}
The achievability scheme for the PIR problem under asymmetric traffic constraints is inspired by the PIR schemes in \cite{JafarPIR,wei2017fundamental}. Our achievable scheme applies message symmetry, and side information exploitation as in \cite{JafarPIR,wei2017fundamental}. However, due to the asymmetric traffic constraints, database symmetry cannot be applied. In an alternative view, we use the side information in an asymmetric fashion among the databases. The most relevant achievable scheme to our achievable scheme here is the scheme in \cite{wei2017fundamental}, in which the bits stored in the user's cache is exploited differently depending on the caching ratio. We begin the discussion by presenting the $M=3$, $N=2$ case as a concrete example to illustrate the main concepts of the scheme.

\subsection{Motivating Example: $M=3$ Messages, $N=2$ Databases} \label{motivating-ex}
In this section, we show the achievability scheme for $M=3$, $N=2$. We first carry out the minimization in \eqref{upper_bound} over $n_1,n_2 \in \{1,2\}$. In this case, we have 4 upper bounds (or effectively 3 bounds if $n_1 \leq n_2$ restriction is applied). By taking the minimum of these bounds for every $\lambda_2 \in [0,1]$, we have the following explicit upper bound on the capacity as a function of $\lambda_2$ (which is in bijection to $\tau_2$)
\begin{align}\label{ub32}
C(\lambda_2) \leq
\left\{
\begin{array}{ll}
\frac{1+3\lambda_2}{3(1+\lambda_2)}, & 0 \leq \lambda_2 \leq \frac{1}{3} \\
\frac{2(1+2\lambda_2)}{5(1+\lambda_2)}, & \frac{1}{3} \leq \lambda_2 \leq \frac{3}{4} \\
\frac{4}{7}, & \frac{3}{4} \leq\lambda_2 \leq 1
\end{array}
\right.
\end{align}

To show the achievability of the upper bound in (\ref{ub32}), let $a_i$, $b_i$, $c_i$ denote randomly and independently permuted symbols of messages $W_1$, $W_2$, $W_3$, respectively. Define $s_2 \in \{0,1,2\}$ to be the number of side information symbols that are used simultaneously in database 2 within the initial round of downloads.  First, we show the achievability of the corner points, i.e., the achievability of the points corresponding to $\lambda_2 \in \{0, \frac{1}{3}, \frac{3}{4}, 1\}$.

\subsubsection{Achievability of the Corner Points}\label{M3N2 corner}
\paragraph{The $\lambda_2=0$ Corner Point:} $\lambda_2=0$ means that the second database does not return any answer strings. The optimal achievable scheme is to download all files from the first database (see Table~\ref{c2=0}). This scheme achieves $R=\frac{1}{3}=C(0)$.

\begin{table}[h]
	\centering
	\caption{The query table for $M=3$, $N=2$, $\lambda_2=0$.}
	\label{c2=0}
	\begin{tabular}{|c|c|}
		\hline
		Database 1 & Database 2 \\
		\hline
		$a_1,b_1,c_1$ &  \\
		\hline
		
	\end{tabular}
\end{table}

\paragraph{The $\lambda_2=1$ Corner Point:} Since $\lambda_1=1$ by definition, $\lambda_2=1$ means that a symmetric scheme can be applied to both databases. Thus, the optimal achievable scheme is the optimal symmetric scheme in \cite{JafarPIR} (see Table~\ref{c2=1}). We present the scheme here for completeness. In this scheme, the user starts with downloading the individual symbols $a_1, b_1, c_1$ from database 1. Since $\lambda_2=1$, database symmetry can be applied, hence the user downloads $a_2, b_2, c_2$ from database 2. Note that in this case, the user does not exploit the side information generated from database 1 in the first round of downloads, but rather downloads new individual symbols, hence $s_2=0$ in this case. The undesired symbols $b_i, c_i$, $i=1,2$ can be exploited in the other database. This can be done by downloading $a_3+b_2$, $a_4+c_2$ from database 1, and similarly by applying database symmetry, the user downloads $a_5+b_1$, $a_6+c_1$ from database 2. In order to satisfy the privacy constraint, the user applies the message symmetry and downloads $b_3+c_3$ from database 1, and $b_4+c_4$ from database 2. Finally, the user exploits the newly generated side information by downloading $a_7+b_4+c_4$ from database 1, and $a_8+b_3+c_3$ from database 2. Consequently, the user downloads $L=8$ symbols in $14$ downloads which results in $R=\frac{8}{14}=\frac{4}{7}=C(1)$.

\begin{table}[h]
	\centering
	\caption{The query table for $M=3$, $N=2$, $\lambda_2=1$.}
	\label{c2=1}
	\begin{tabular}{|c|c|}
		\hline
		Database 1 & Database 2 \\
		\hline
		$a_1,b_1,c_1$ & $a_2,b_2,c_2$ \\
		\hline
		$a_3+b_2$ & $a_5+b_1$ \\
		$a_4+c_2$ & $a_6+c_1$ \\
		$b_3+c_3$ & $b_4+c_4$ \\
		\hline
		$a_7+b_4+c_4$ & $a_8+b_3+c_3$ \\
		\hline
	\end{tabular}
\end{table}

\paragraph{The $\lambda_2=\frac{3}{4}$ Corner Point:} The user can cut the first round of downloads in database 2 and exploit the side information generated from database 1 directly in the form of sums of 2, i.e., the user downloads $a_1, b_1, c_1$ from database 1 and then exploits the undesired symbols as side information by downloading $a_2+b_1$, $a_3+c_1$ from database 2. The user then applies message symmetry and downloads $b_2+c_2$. Since the user uses 1 bit of side information in the initial download round from database 2, $s_2=1$ in this case. Finally, the user exploits the undesired sum $b_2+c_2$ from database 2 as a side information in database 1 and downloads $a_4+b_2+c_2$. Using this scheme the user downloads $4$ symbols from database 1 and $3$ symbols from database 2, hence $\lambda_2=\frac{3}{4}$. The user downloads $L=4$ desired symbols out of $7$ downloads, thus $R=\frac{4}{7}=C(\frac{3}{4})$. The privacy is satisfied since $W_1$, $W_2$, $W_3$ are independently and randomly permuted, and since the scheme includes all the possible combinations of the sums in any round. The query table for this scheme is given in Table~\ref{c2=3/4}. We note that this scheme is exactly the asymmetric achievable scheme presented in \cite{arbmsgPIR}.

\begin{table}[h]
	\centering
	\caption{The query table for $M=3$, $N=2$, $\lambda_2=\frac{3}{4}$.}
	\label{c2=3/4}
	\begin{tabular}{|c|c|}
		\hline
		Database 1 & Database 2 \\
		\hline
		$a_1,b_1,c_1$ & \\
		\hline
		 & $a_2+b_1$ \\
		 & $a_3+c_1$ \\
		 & $b_2+c_2$ \\
		\hline
		$a_4+b_2+c_2$ & \\
		\hline
	\end{tabular}
\end{table}

\paragraph{The $\lambda_2=\frac{1}{3}$ Corner Point:} In this case, the user downloads $a_1,b_1,c_1$ from database 1. In database 2, the user exploits the side information $b_1,c_1$ simultaneously and downloads $a_2+b_1+c_1$. Due to the fact that 2 side information symbols are used simultaneously in the initial round of download from database 2, $s_2=2$ in this case. Using this scheme the user downloads $3$ symbols from database 1 and $1$ symbol from database 2, therefore $\lambda_2=\frac{1}{3}$. The user downloads $L=2$ desired symbols in $4$ downloads, hence $R=\frac{1}{2}=C(\frac{1}{3})$. The privacy follows by the same argument as in the previous case. The query table for this case is given in Table~\ref{c2=1/3}.

\begin{table}[h]
	\centering
	\caption{The query table for $M=3$, $N=2$, $\lambda_2=\frac{1}{3}$.}
	\label{c2=1/3}
	\begin{tabular}{|c|c|}
		\hline
		Database 1 & Database 2 \\
		\hline
		$a_1,b_1,c_1$ & \\
		\hline
		&  \\
		\hline
		& $a_2+b_1+c_1$\\
		\hline
	\end{tabular}
\end{table}

\subsubsection{Achievability of Non-Corner Points}\label{M3N2 non-corner}
In the following, we show that by combining the achievable schemes of the corner points over different symbols, the upper bound in \eqref{ub32} is tight for any $\lambda_2$. We note that the privacy constraint is still satisfied after this combination as each scheme operates over different sets of symbols and the fact that each scheme satisfies the privacy constraint individually. A formal argument for proving that combination of private schemes remains private can be found in \cite[Theorem~4]{arbmsgPIR}. Let $\nu_{s_2}$, where $s_2=0,1,2$, denote the number of repetitions of the scheme that uses $s_2$ side information symbols simultaneously in the first round of download in database 2. By convention, let $\nu_3$ denote the number of repetitions of the trivial retrieval scheme, i.e., when the retrieval is solely done from database 1.

\paragraph{The $0 \leq \lambda_2 \leq \frac{1}{3}$ Regime:} We combine the achievable scheme of $\lambda_2=0$ corner point with the achievable scheme of $\lambda_2=\frac{1}{3}$ corner point. The achievable scheme of $\lambda_2=0$ downloads 3 symbols from database 1 and 0 symbols from database 2. We perform this scheme $\nu_3$ repetitions. The achievable scheme of $\lambda_2=\frac{1}{3}$ downloads 3 symbols from database 1 and 1 symbol from database 2. We perform this scheme $\nu_2$ repetitions. Under the asymmetric traffic constraints, this results in the following system of equations
\begin{align}
	3\nu_3+3\nu_2&=t_1 \\
	      \nu_2&=\lambda_2 t_1
\end{align}
This system has a unique solution (parametrized by $t_1$) of $\nu_2=\lambda_2 t_1$ and $\nu_3=\frac{1-3\lambda_2}{3} t_1$. Note that $\nu_3 \geq 0$ in the regime of $0 \leq \lambda_2 \leq \frac{1}{3}$. Since the scheme of $\lambda_2=0$ downloads 1 symbol from the desired message and the scheme of $\lambda_2=\frac{1}{3}$ downloads 2 symbols from the desired message. The achievable rate $R(\lambda_2)$ is given by
\begin{align}
R(\lambda_2)=\frac{2\nu_2+\nu_3}{(1+\lambda_2)t_1}=\frac{1+3\lambda_2}{3(1+\lambda_2)}=C(\lambda_2), \qquad 0 \leq \lambda_2 \leq \frac{1}{3}
\end{align}

\paragraph{The $\frac{1}{3} \leq \lambda_2 \leq \frac{3}{4}$ Regime:} Similarly, the user combines the achievable schemes of $\lambda_2=\frac{1}{3}$ and $\lambda_2=\frac{3}{4}$ corner points. The user applies the scheme of $\lambda_2=\frac{1}{3}$ for $\nu_2$ repetitions, which downloads 3 symbols from database 1 and 1 symbol from database 2 and has $L=2$. The user applies the scheme of $\lambda_2=\frac{3}{4}$ for $\nu_1$ repetitions, which downloads 4 symbols from database 1 and 3 symbols from database 2 and has $L=4$. This results in the following system of equations
\begin{align}
4\nu_1+3\nu_2&=t_1 \\
3\nu_1+\nu_2&=\lambda_2 t_1
\end{align}
which has the following solution: $\nu_1=\frac{-1+3\lambda_2}{5}t_1 \geq 0$ and $\nu_2=\frac{3-4\lambda_2}{5}t_1 \geq 0$ in the regime of $\frac{1}{3} \leq \lambda_2 \leq \frac{3}{4}$. Consequently, the achievable rate is given by
\begin{align}
R(\lambda_2)=\frac{4\nu_1+2\nu_2}{(1+\lambda_2)t_1}=\frac{2(1+2\lambda_2)}{5(1+\lambda_2)}=C(\lambda_2), \qquad \frac{1}{3} \leq \lambda_2 \leq \frac{3}{4}
\end{align}

\paragraph{The $\frac{3}{4} \leq \lambda_2 \leq 1$ Regime:} The user combines the  achievable schemes of $\lambda_2=\frac{3}{4}$ and $\lambda_2=1$ corner points. The user repeats the scheme of $\lambda_2=\frac{3}{4}$ for $\nu_1$ repetitions, and the scheme of $\lambda_2=1$ for $\nu_0$ repetitions. This results in the following system of equations
\begin{align}
4\nu_1+7\nu_0&=t_1 \\
3\nu_1+7\nu_0&=\lambda_2 t_1
\end{align}
The solution for this system is given by: $\nu_1=(1-\lambda_2)t_1 \geq 0$ and $\nu_0=\frac{-3+4\lambda_2}{7} t_1 \geq 0$ in the regime of $\frac{3}{4} \leq \lambda_2 \leq 1$. The corresponding rate is given by
\begin{align}
R(\lambda_2)=\frac{4\nu_1+8\nu_0}{(1+\lambda_2)t_1}=\frac{4}{7}=C(\lambda_2), \qquad \frac{3}{4} \leq \lambda_2 \leq 1
\end{align}

\paragraph{Specific Example for Non-Corner Points, $\lambda_2=\frac{1}{2}$:} The query table for this case is given in Table~\ref{c2=1/2}. The user applies the scheme of $\lambda_2=\frac{3}{4}$ for $\nu_1=\frac{-1+3\lambda_2}{5}t_1=\frac{1}{10}t_1$ repetitions, and the scheme of $\lambda_2=\frac{1}{3}$ for $\nu_2=\frac{3-4\lambda_2}{5}t_1=\frac{1}{5}t_1$ repetitions. Choosing $t_1=10$, we have $\nu_1=1$ and $\nu_2=2$. The scheme downloads 10 symbols from database 1 and 5 symbols from database 2, thus, $\lambda_2=\frac{1}{2}$. The scheme downloads $8$ symbols in $15$ downloads, hence $R(\frac{1}{2})=\frac{8}{15}=\frac{2(1+2\lambda_2)}{5(1+\lambda_2)}=C(\frac{1}{2})$.

\begin{table}[h]
	\centering
	\caption{The query table for $M=3$, $N=2$, $\lambda_2=\frac{1}{2}$.}
	\label{c2=1/2}
	\begin{tabular}{|c|c|}
		\hline
		Database 1 & Database 2 \\
		\hline
		$a_1,b_1,c_1$ & $a_2+b_1$ \\
					  & $a_3+c_1$ \\
					  & $b_2+c_2$ \\
		\hline
		$a_4+b_2+c_2$ &           \\
		\hline\hline
		$a_5, b_3, c_3$ & $a_6+b_3+c_3$ \\
		\hline\hline
		$a_7, b_4, c_4$ & $a_8+b_4+c_4$ \\
		\hline
	\end{tabular}
\end{table}


\subsection{Description of the General Scheme}
In this section, we describe the general achievable scheme that achieves the retrieval rates in Theorem~\ref{Thm2}. We first show explicitly the achievability schemes for corner points, i.e., the achievability scheme for every monotone non-decreasing sequence $\{n_i\}_{i=0}^{M-1} \subset \{1, \cdots, N\}^M$. We note that our achievability scheme is different in two key steps: First regarding the database symmetry, we note that it is not applied over all databases directly as in \cite{JafarPIR}, but rather it is applied over groups of databases, such as, group 0 includes databases 1 through $n_0$, group 1 includes databases $n_0+1$ through $n_1$, etc. Second, regarding the exploitation of side information step, we note that each group of databases exploits side information differently in the \emph{initial} round of downloading. More specifically, we note that group 0 of databases do not exploit any side information in the initial round of the download, group 1 exploits 1 side information symbol in the initial round of the download, group 2 exploits sums of 2 side information symbols in the initial round of the download, and so on.

Next, we show that for non-corner points, time-sharing between corner points is achievable and this concludes the achievability proof of Theorem~\ref{Thm2}.

\subsubsection{Achievability Scheme for the Corner Points}
Let $s_n \in \{0, 1, \cdots, M-1\}$ denote the number of side information symbols that are used simultaneously in the initial round of downloads at the $n$th database. For a given non-decreasing sequence $\{n_i\}_{i=0}^{M-1} \subset \{1, \cdots, N\}^M$, let $s_n=i$ for all $n_{i-1}+1 \leq n \leq n_i$ with $n_{-1}=0$ by convention. Denote $\cs=\{i: s_n=i \:\text{for some}\: n \in \{1, \cdots, N\} \}$. We follow the round and stage definitions in \cite{MPIRjournal}. The $k$th round is the download queries that admit a sum of $k$ different messages ($k$-sum in \cite{JafarPIR}). A stage of the $k$th round is a query block of the $k$th round that exhausts all $\binom{M}{k}$ combinations of the $k$-sum. Denote $y_\ell[k]$ to be the number of stages in round $k$ downloaded from the $n$th database, such that $n_{\ell-1}+1 \leq n \leq n_\ell$. The details of the achievable scheme are as follows:

\begin{enumerate}
	\item \emph{Initialization:} The user permutes each message independently and uniformly using a random interleaver, i.e.,
	\begin{align}
	x_m(i)=W_m(\pi_m(i)), \quad i \in \{1, \cdots, L\}
	\end{align}
	where $x_m(i)$ is the $i$th symbol of the permuted $W_m$, $\pi_m(\cdot)$ is a random interleaver for the $m$th message that is chosen independently, uniformly, and privately at the user's side. From the $n$th database where $1 \leq n \leq n_0$, the user downloads $\prod_{s \in \cs} \binom{M-2}{s-1}$ symbols from the desired message. The user sets the round index $k=1$. I.e., the user starts downloading the desired symbols from $y_0[1]=\prod_{s \in \cs} \binom{M-2}{s-1}$ different stages.
	
	\item \emph{Message symmetry:} To satisfy the privacy constraint, for each stage initiated in the previous step, the user completes the stage by downloading the remaining $\binom{M-1}{k-1}$ $k$-sum combinations that do not include the desired symbols, in particular, if $k=1$, the user downloads $\prod_{s \in \cs} \binom{M-2}{s-1}$ individual symbols from each undesired message.
	
	\item \emph{Database symmetry:} Due to the asymmetric traffic constraints, the original database symmetry step in \cite{JafarPIR} cannot be applied directly to our problem. Instead, we divide the databases into groups. Group $\ell \in \cs$ corresponds to databases $n_{\ell-1}+1$ to $n_{\ell}$. Database symmetry is applied within each group only. Consequently, the user repeats step~2 over each group of databases, in particular, if $k=1$, the user downloads $\prod_{s \in \cs} \binom{M-2}{s-1}$ individual symbols from each message from the first $n_0$ databases (group 1).
	
	\item \emph{Exploitation of side information:} This step is also different from \cite{JafarPIR} because of the asymmetric traffic constraints. In order to create different lengths of the answer strings, the initial exploitation of side information is group-dependent as well. More specifically, the undesired symbols  downloaded within the $k$th round (the $k$-sums that do not include the desired message) are used as side information in the $(k+1)$th round. This exploitation of side information is performed by downloading $(k+1)$-sum consisting of 1 desired symbol and a $k$-sum of undesired symbols only that were generated in the $k$th round. However, the main difference from \cite{JafarPIR} is that for the $n$th database, if $s_n>k$, then this database does not exploit the side information generated in the $k$th round. Thus, the $n$th database belonging to the $\ell$th group exploits the side information generated in the $k$th round from all databases except itself if $s_n \leq k$. Moreover, for $s_n=k$, extra side information can be used in the $n$th database. This is because the user can form $n_0\prod_{s \in \cs\setminus \{s_n\}} \binom{M-2}{s-1}$ extra stages of side information by constructing $k$-sums of the undesired symbols in round 1 from the databases in group 0.
	
	\item \emph{Repeat} steps 2, 3, 4 after setting $k=k+1$ until $k=M$.
	
	\item \emph{Shuffling the order of the queries:} By shuffling the order of the queries uniformly, all possible queries can be made equally likely regardless of the message index. This guarantees the privacy.
\end{enumerate}

\subsubsection{Achievability Scheme for Non-Corner Points}
In this section, we show that achievability schemes for non-corner points can be derived by time-sharing between the nearest corner points, i.e., the achievable scheme under $\bt$ constraints is performed by time-sharing between the corner points of an $N$-dimensional polytope that enclose the traffic vector $\bt$. The following lemma formalizes the time-sharing argument. Lemma~\ref{time-sharing} can be thought of as an adaptation of \cite[Theorem~4]{arbmsgPIR} and \cite[Lemma~1]{tandon2017capacity} to the PIR problem under asymmetric traffic constraints.

\begin{lemma}[Time-sharing]\label{time-sharing}
	For the PIR problem under asymmetric traffic constraints $\bt$, let the retrieval rate $R(\bt_i)$ be achievable for the traffic ratio vector $\bt_i$ for all $i \in \{1,\cdots, N\}$. Moreover, assume that $\bt=\sum_{i=1}^{N} \alpha_i \bt_i$ for some $\{\alpha_i\}_{i=1}^N$ such that $\alpha_i \geq 0$, for all $i$, and  $\sum_{i=1}^N \alpha_i=1$. Then, the following retrieval rate $R(\bt)$ is achievable,
	\begin{align}
	R(\bt)=\sum_{i=1}^{N} \alpha_i R(\bt_i)
	\end{align}
\end{lemma}

\begin{Proof}
	Let $\text{PIR}_i$ denote the PIR scheme that achieves retrieval rate $R(\bt_i)$ for a traffic ratio vector $\bt_i$. Denote the total download of $\text{PIR}_i$ by $D_i$ and the corresponding message length by $L_i$.
	
	Now, construct the following PIR scheme with total download $D$ and message length $L$. For each database, concatenate the queries from the $N$ PIR schemes with ensuring that each symbol is queried by one PIR scheme only. Hence, $D=\sum_{i=1}^{N} D_i$, such that $D_i=\alpha_i D$, for $i \in \{1,\cdots,N\}$, and the download from the $n$th database is $t_n(\bt)=\sum_{i=1}^{N} t_n(\bt_i)$. This concatenation of the achievable schemes is feasible under asymmetric traffic constraints since $\bt=\sum_{i=1}^{N} \alpha_i \bt_i$. To see this, we note that the $n$th element of the traffic ratio vector $\tau_n$ is given by
	\begin{align}
	\tau_n=\frac{t_n(\bt)}{D}=\frac{\sum_{i=1}^N t_n(\bt_i)}{D}=\frac{\sum_{i=1}^N \tau_n^{(i)} D_i}{D}=\frac{\sum_{i=1}^N \tau_n^{(i)} \alpha_i D}{D}=\sum_{i=1}^{N} \alpha_i \tau_n^{(i)}
	\end{align}
	where $\tau_n^{(i)}$ denotes the $n$th element in $\bt_i$. Since these implications are true for each element in $\bt$, we have $\bt=\sum_{i=1}^{N} \alpha_i \bt_i$ as required.
	
	$\text{PIR}_i$ scheme downloads $L_i$ symbols from the desired messages, such that
	\begin{align}
	L_i=R(\bt_i) D_i=\alpha_i R(\bt_i) D
	\end{align}
	Hence, the total message length by concatenating all the achievable schemes together is
	\begin{align}
	L=\sum_{i=1}^{N} L_i=\sum_{i=1}^{N} \alpha_i R(\bt_i) D
	\end{align}
	and the corresponding achievable rate is given by
	\begin{align}
	R(\bt)=\frac{L}{D}=\sum_{i=1}^{N} \alpha_i R(\bt_i)
	\end{align}
	
	The reliability constraint follows from the reliability of each PIR scheme. The privacy constraint is satisfied due to the fact that each PIR scheme operates on a different portion of the messages and these portions are picked uniformly and independently. Hence, the privacy constraint for the concatenated scheme follows from the privacy of each PIR scheme. A formal treatment of the privacy constraint of concatenated schemes can be found in \cite{arbmsgPIR}.
\end{Proof}

Thus, Lemma~\ref{time-sharing} provides an achievability proof for any traffic ratio vector $\bt$ that is not a corner point. Finally, we have the following remark regarding this time-sharing lemma.

\begin{remark}
	We note that although the vector $\boldsymbol{\lambda}=(\lambda_1, \cdots, \lambda_N)$ is in bijection with $\bt=(\tau_1, \cdots, \tau_N)$, the time-sharing argument in Lemma~\ref{time-sharing} does not hold for $R(\boldsymbol{\lambda})$. This is due to the fact that $R(\boldsymbol{\lambda})$ is a non-linear function of $\boldsymbol{\lambda}$ whereas $R(\bt)$ is an affine function of $\bt$.
\end{remark}

\subsection{Decodability, Privacy, and Calculation of the Achievable Rate}
In this section, we prove the decodability, privacy and the achievable rate in Theorem~\ref{Thm2}. We note that it suffices to consider the corner points only, as Lemma~\ref{time-sharing} settles the decodability, privacy and achievable rate for non-corner points based on the existence of feasible PIR schemes that achieve the corner points.

\paragraph{Decodability:} By construction, in the $(k+1)$th round at the $n$th database, the user exploits the side information generated in the $k$th round in the remaining active databases by adding 1 symbol of the desired message with $(k-1)$-sum of undesired messages which was downloaded previously in the $k$th round. Moreover, for the $n$th database belonging to the $\ell$th group at the $(\ell+1)$th round, the user adds every $\ell$ symbols of the undesired symbols downloaded from group 0 to make one side information symbol. Since the user downloads $\prod_{\ell \in \cs} \binom{M-2}{\ell-1}$ symbols from every database in the first $n_0$ databases (group 0), the user can exploit such side information to initiate $n_0 \prod_{\ell \in \cs \setminus \{\ell\}} \binom{M-2}{\ell-1}$ stages in the $(\ell+1)$th round from every database in group $\ell$. Since all side information symbols used in the $(k+1)$th round are decodable in the $k$th round or from round 1, the user cancels out these side information symbols and is left with symbols from the desired message.

\paragraph{Privacy:} For every stage of the $k$th round initiated in the exploitation of the side information step, the user completes the stage by including all the remaining $\binom{M-1}{k-1}$ undesired symbols. This implies that all $\binom{M}{k}$ combinations of the $k$-sum are included at each round. Thus, the structure of the queries is the same for any desired message. The privacy constraint in \eqref{privacy_constraint} is satisfied by the random and independent permutation of each message and the random shuffling of the order of the queries. This ensures that all queries are equally likely independent of the desired message index.

\paragraph{Calculation of the Achievable Rate:} For a corner point characterized by the non-decreasing sequence $\{n_i\}_{i=0}^{M-1}$, as mentioned before, we denote $y_\ell[k]$ to be the number of stages that admit $k$-sums downloaded from any database belonging to the $\ell$th group, i.e., the $n$th database such that $n_{\ell-1}+1 \leq n \leq n_\ell$. By construction, we observe that all databases belonging to the $\ell$th group are inactive until the $(\ell+1)$th round as the user initiates download in such databases by exploiting $\ell$ bits of side information simultaneously by definition of the group. Consequently, we have the initial condition $y_\ell[k]=0$ for $k \leq \ell$. Since the user downloads $\prod_{s \in \cs} \binom{M-2}{s-1}$ individual symbols (i.e., from round 1) from group 0, we have the following initial condition $y_0[1]=\prod_{s \in \cs} \binom{M-2}{s-1}$.

Now, we note from the side information exploitation step that the user initiates new stages in the $k$th round from the $n$th database depending on the number of stages of the $(k-1)$th round for group 0 and group 1 (i.e., for $1 \leq n \leq n_1$). More specifically, for the $n$th database belonging to group 0, the user considers all the undesired symbols downloaded from all databases (except the $n$th database) in the $(k-1)$th round as side information. Since database symmetry applies over each group, and from the fact that each stage in the $(k-1)$th round initiates a stage in the $k$th round, we have
\begin{align}
y_0[k]=(n_0-1)y_0[k-1]+\sum_{\ell \in \cs\setminus \{0\}}(n_\ell-n_{\ell-1}) y_\ell[k-1]
\end{align}
where the left side is the total number of stages in the $(k-1)$th round from all the $N-1$ databases (i.e., except for the $n$th database that belongs to group 0). The same argument holds for group 1 as well, hence
\begin{align}
y_1[k]=(n_1-n_0-1)y_1[k-1]+\sum_{\ell \in \cs\setminus \{1\}}(n_\ell-n_{\ell-1}) y_\ell[k-1]
\end{align}
where $(n_1-n_0-1)$ denotes the number of databases in group 1 other than the $n$th database.

For a database belonging to the $\ell$th group such that $\ell \geq 2$, the user can generate extra stages by exploiting the symbols downloaded in round 1. To initiate one stage in the $(\ell+1)$th round, the user needs to combine symbols from $\frac{\binom{M-1}{\ell} \ell}{M-1}=\binom{M-2}{\ell-1}$ stages. Therefore, the number of stages initiated in the $(\ell+1)$th round as a consequence of the side information in round 1 is $\xi_\ell=\frac{y_0[1]}{\binom{M-2}{\ell-1}}=\prod_{s \in \cs \setminus \{\ell\}} \binom{M-2}{s-1}$. Since these extra side information can be used once (at the $(\ell+1)$th round only) and after that for the $k$th round, the database exploits the side information generated in the $(k-1)$th round only. We represent this one-time exploitation of side information in the $(\ell+1)$th round by the Kronecker delta function $\delta[k-\ell-1]$. Consequently, the number of stages for the $\ell$th group, $\ell \geq 2$ is related via the following difference equation:
\begin{align}
y_\ell[k]=n_0 \xi_\ell \delta[k\!-\!\ell\!-\!1]+(n_\ell\!-\!n_{\ell-1}\!-\!1) y_\ell[k-1]+\sum_{j \in \cs \setminus \{\ell\}} (n_j\!-\!n_{j-1})y_j[k\!-\!1]
\end{align}

Now, we are ready to characterize $\bt(\mathbf{n})$ and $R(\bt(\mathbf{n}))$ in terms of $y_\ell[k]$, where $\ell \in \cs$ and $k=1, \cdots, M$. For any stage in the $k$th round, the user downloads $\binom{M-1}{k-1}$ desired symbols from a total of $\binom{M}{k}$ downloads. Therefore, from a database belonging to the $\ell$th group, the user downloads $\sum_{k=1}^{M}\binom{M-1}{k-1} y_\ell[k]$ desired symbols from a total of $\sum_{k=1}^{M}\binom{M}{k} y_\ell[k]$. The number of databases belonging to the $\ell$th group is given by $n_\ell-n_{\ell-1}$. Therefore, the total download is given by,
\begin{align}
\sum_{n=1}^{N} t_n(\bt(\mathbf{n}))=\sum_{\ell \in \cs} \sum_{k=1}^{M} \binom{M}{k} y_\ell[k] (n_\ell-n_{\ell-1})
\end{align}

Thus, the traffic ratio of the $n$th database belonging to the $\ell$th group (i.e., $n_{\ell-1}+1 \leq n \leq n_\ell$) corresponding to $\mathbf{n}=\{n_i\}_{i=0}^{M-1}$ is given by
\begin{align}
	\tau_n(\mathbf{n})=\tilde{\tau}_\ell=\frac{\sum_{k=1}^M \binom{M}{k} y_\ell[k]}{\sum_{\ell \in \cs} \sum_{k=1}^M \binom{M}{k} y_\ell[k] (n_\ell-n_{\ell-1})},\quad n_{\ell-1}+1 \leq n \leq n_\ell
\end{align}
Furthermore, the total desired symbols from all databases is given by
\begin{align}
L(\bt(\mathbf{n}))=\sum_{\ell \in \cs} \sum_{k=1}^{M}\binom{M-1}{k-1} y_\ell[k](n_\ell-n_{\ell-1})
\end{align}
which further leads to the following achievable rate
\begin{align}
	R(\bt(\mathbf{n}))=\frac{\sum_{\ell \in \cs} \sum_{k=1}^M \binom{M-1}{k-1} y_\ell[k] (n_\ell-n_{\ell-1})}{\sum_{\ell \in \cs} \sum_{k=1}^M \binom{M}{k} y_\ell[k] (n_\ell-n_{\ell-1})}
\end{align}

\section{Optimality of $M=2$ and $M=3$ Cases}\label{optimal_achievability}
In this section, we prove Corollary~\ref{optimality23}, i.e., we prove that the capacity of the PIR problem under asymmetric traffic constraints $C(\bt)$ for $M=2,3$ is given by \eqref{capacityM32}. We note that since the upper bound in Theorem~\ref{Thm1} is affine in $\bt$ and time-sharing rates are achievable from Lemma~\ref{time-sharing}, it suffices to prove the optimality of all corner points to settle the PIR capacity $C(\bt)$ for $M=2,3$. In the following, we use Theorem~\ref{Thm1} and Theorem~\ref{Thm2} to show the optimality of these corner points.

\subsection{$M=2$ Messages}
We start the proof from the achievability side. From Theorem~\ref{Thm2}, the corner points are specified by the non-decreasing sequence $\mathbf{n}=(n_0,n_1)$. In this case, the system of difference equations in \eqref{difference_eqn} is reduced to
\begin{align}
y_0[k]&=(n_0-1)y_0[k-1] \\
y_1[k]&=n_0 y_0[k-1]
\end{align}
for $k=1,2$, where $y_0[1]=1$, and $y_1[1]=0$. Hence, $y_0[2]=n_0-1$, and $y_1[2]=n_0$. Hence, the total downloads for the corner point $\mathbf{n}=(n_0,n_1)$ is
\begin{align}
\sum_{n=1}^{N} t_n(\bt(\mathbf{n}))=\sum_{\ell=0}^{1}\sum_{k=1}^{2} \binom{2}{k} y_\ell[k](n_\ell-n_{\ell-1})=n_0(n_1+1)
\end{align}
Thus, from Theorem~\ref{Thm2}, the traffic-ratio vector $\bt(\mathbf{n})$ is given by
\begin{align}
\tilde{\tau}_0&=\frac{\binom{2}{1}y_0[1]+\binom{2}{2} y_0[2]}{\sum_{n=1}^{N} t_n(\bt(\mathbf{n}))}=\frac{n_0+1}{n_0(n_1+1)}\\
\tilde{\tau}_1&=\frac{\binom{2}{1}y_1[1]+\binom{2}{2} y_1[2]}{\sum_{n=1}^{N} t_n(\bt(\mathbf{n}))}=\frac{1}{n_1+1}
\end{align}
where $\tau_n=\tilde{\tau}_0$, for $1 \leq n\leq n_0$, and $\tau_n=\tilde{\tau}_1$, for $n_0+1\leq n \leq n_1$, and $\tau_n=0$ otherwise.
For the desired symbols, the user downloads $L_0(\bt(\mathbf{n}))$ symbols from the $n$th database when $1 \leq n \leq n_0$, and $L_1(\bt(\mathbf{n}))$ symbols from the $n$th database when $n_0+1 \leq n \leq n_1$
\begin{align}
L_0(\bt(\mathbf{n}))=y_0[1]+y_0[2]=n_0 \label{water-filling2a} \\
L_1(\bt(\mathbf{n}))=y_1[1]+y_1[2]=n_0 \label{water-filling2b}
\end{align}
Consequently, $L=n_0 L_0+(n_1-n_0)L_1=n_0n_1$, and the achievable retrieval rate $R(\bt(\mathbf{n}))$ is given by
\begin{align}
R(\bt(\mathbf{n}))=\frac{L(\bt(\mathbf{n}))}{\sum_{n=1}^{N} t_n(\bt(\mathbf{n}))}=\frac{n_1}{n_1+1}
\end{align}

For the converse, we evaluate the bound in \eqref{upper_bound} (without the minimization) for $n_1=n_0$, i.e., we substitute with $n_0$ in the argument of the upper bound. Then, we have the following upper bound
\begin{align}
R(\bt(\mathbf{n})) &\leq \frac{1+\frac{\sum_{n=n_0+1}^{N} \tau_n}{n_0}}{1+\frac{1}{n_0}}\\
                   &=\frac{1+\frac{(n_1-n_0)\tilde{\tau}_1}{n_0}}{1+\frac{1}{n_0}} \\
                   &=\frac{n_1}{n_1+1}
\end{align}
This concludes the optimality of our achievable scheme for $M=2$.

\subsection{$M=3$ Messages}
Similarly, for the corner point specified by the non-decreasing sequence $\mathbf{n}=(n_0,n_1,n_2)$, we have the following system of difference equations for $k=1,2,3$
\begin{align}
y_0[k]&=(n_0-1)y_0[k-1]+(n_1-n_0)y_1[k-1]+(n_2-n_1)y_2[k-1] \\
y_1[k]&=n_0y_0[k-1]+(n_1-n_0-1)y_1[k-1]+(n_2-n_1)y_2[k-1] \\
y_2[k]&=n_0\delta[k-3]+n_0y_0[k-1]+(n_1-n_0)y_1[k-1]+(n_2-n_1-1)y_2[k-1]
\end{align}
with the initial conditions $y_0[1]=1$, $y_1[1]=0$, and $y_2[1]=y_2[2]=0$. Evaluating $y_\ell[k]$, for $\ell=0,1,2$, and $k=1,2,3$ recursively leads to $y_0[2]=n_0-1$, $y_1[2]=n_0$, $y_0[3]=n_1n_0-2n_0+1$, $y_1[3]=n_1n_0-2n_0$, and $y_2[3]=n_1n_0$. This leads to the following total download
\begin{align}
\sum_{n=1}^{N} t_n(\bt(\mathbf{n}))=\sum_{\ell=0}^{2} \sum_{k=1}^{3} \binom{3}{k} y_\ell[k](n_\ell-n_{\ell-1})=n_0(n_1n_2+n_1+1)
\end{align}
The sequence $\mathbf{n}=(n_0,n_1,n_2)$ specifies the traffic ratio vector $\bt(\mathbf{n})$ such that
\begin{align}
\tilde{\tau}_0&=\frac{n_0n_1+n_0+1}{n_0(n_2n_1+n_1+1)}\\
\tilde{\tau}_1&=\frac{n_1+1}{n_2n_1+n_1+1}\\
\tilde{\tau}_2&=\frac{n_1}{n_2n_1+n_1+1}
\end{align}
where $\tau_n=\tilde{\tau}_0$ for $1 \leq n \leq n_0$, $\tau_n=\tilde{\tau}_1$ for $n_0+1\leq n \leq n_1$, $\tau=\tilde{\tau}_2$ for $n_1+1 \leq n \leq n_2$, and $\tau_n=0$ otherwise.

For the desired symbols, the user downloads $L_0(\bt(\mathbf{n}))$ symbols from the $n$th database if $1 \leq n \leq n_0$, $L_1(\bt(\mathbf{n}))$ symbols if $n_0+1 \leq n \leq n_1$, and $L_2(\bt(\mathbf{n}))$ symbols if $n_1+1 \leq n \leq n_2$, hence
\begin{align}\label{water-filling3}
L_\ell(\bt(\mathbf{n}))=\sum_{k=1}^{3} \binom{2}{k-1} y_\ell[k]=n_0 n_1, \quad \ell=0,1,2
\end{align}
Consequently, the following rate is achievable
\begin{align}
R(\bt(\mathbf{n}))=\frac{n_1 n_2}{n_1n_2+n_1+1}
\end{align}

For the converse, pick $(n_1,n_2)$ in the converse bound to be $(n_0,n_1)$, which leads to the following bound
\begin{align}
R(\bt(\mathbf{n})) &\leq \frac{1+\frac{\sum_{n=n_0+1}^{N} \tau_n}{n_0}+\frac{\sum_{n=n_1+1}^{N} \tau_n}{n_0n_1}}{1+\frac{1}{n_0}+\frac{1}{n_0n_1}}\\
&=\frac{1+\frac{(n_1-n_0)\tilde{\tau}_1}{n_0}+\frac{(n_2-n_1)\tilde{\tau}_2}{n_0}+\frac{(n_2-n_1)\tilde{\tau}_2}{n_0n_1}}{1+\frac{1}{n_0}+\frac{1}{n_0n_1}} \\
&=\frac{n_1n_2}{n_1n_2+n_1+1}
\end{align}
This concludes the optimality of our achievable scheme for $M=3$.

\begin{remark}
	We note that, surprisingly, for the corner points of the cases $M=2$ and $M=3$, the number of desired symbols downloaded from each active database is the same irrespective to the traffic ratio of the database; see \eqref{water-filling2a}-\eqref{water-filling2b} for $M=2$ and \eqref{water-filling3} for $M=3$. This suggests that at these corner points, the optimal scheme performs \emph{combinatorial water-filling} for the undesired symbols first, i.e., the $n$th active database downloads $t_n-n_0$ undesired symbols for $M=2$ and $t_n-n_0n_1$ undesired symbols for $M=3$, and then downloads the same number of desired symbols from all active databases.
\end{remark}

\section{Achievable Tradeoff for $N=2$ and Arbitrary $M$}\label{Tradeoff N=2}
For the special case of $N=2$, and an arbitrary $M$, the retrieval rate calculation in Theorem~\ref{Thm2} is significantly simplified. Let $s_2 \in \{0, \cdots, M-1\}$ be the number of side information symbols that are used simultaneously in the initial round of download at the second database. Note that there is a bijection between $s_2$ and the non-decreasing sequence $\mathbf{n}$ as $n_0=n_1=\cdots=n_{s_2-1}=1$, and $n_{s_2}=2$ for any corner point other than the corner point corresponding to the trivial scheme of downloading the contents of the first database.

The user starts with downloading $\binom{M-2}{s_2-1}$ stages of individual symbols (i.e., the user downloads $M\binom{M-2}{s_2-1}$ symbols in round 1 from all messages) from the first database to create 1 stage in the $(s_2+1)$th round. After the initial exploitation of side information, the two databases exchange side information. More specifically, from database 1 in the $(s_2+2k)$th round, where $k=1, \cdots, \left\lfloor \frac{M-s_2}{2}\right\rfloor$, the user exploits the side information generated in database 2 in the $(s_2+2k-1)$th round to download $\binom{M-1}{s_2+2k-1}$ desired symbols (by adding one symbol of the desired symbols to the $(s_2+2k-1)$-sum of undesired symbols generated in database 2) from total download in the $(s_2+2k)$th round of $\binom{M}{s_2+2k}$. Similarly from database 2, in the $(s_2+2k+1)$th round, where $k=0, \cdots, \left\lfloor \frac{M-s_2-1}{2}\right\rfloor$, the user exploits the side information generated in database 1 in the $(s_2+2k)$th round, and downloads $\binom{M-1}{s_2+2k}$ desired symbols from total of $\binom{M}{s_2+2k+1}$ downloads in the $(s_2+2k+1)$th round.

Consequently, we have
\begin{align}
t_1(s_2)&=M\binom{M-2}{s_2-1}+\sum_{k=1}^{\left\lfloor \frac{M-s_2}{2}\right\rfloor} \binom{M}{s_2+2k} \\
t_2(s_2)&=\sum_{k=0}^{\left\lfloor \frac{M-s_2-1}{2}\right\rfloor} \binom{M}{s_2+2k+1}
\end{align}
which further leads to the following total download
\begin{align}
t_1(s_2)+t_2(s_2)=M\binom{M-2}{s_2-1}+\sum_{k=0}^{M-s_2-1} \binom{M}{s_2+k+1}
\end{align}
Thus, the traffic ratio $\tau_2(s_2)$ is given by
\begin{align}
\tau_2(s_2)=\frac{t_2(s_2)}{t_1(s_2)+t_2(s_2)}=\frac{\sum_{k=0}^{\left\lfloor \frac{M-s_2-1}{2}\right\rfloor} \binom{M}{s_2+2k+1}}{M\binom{M-2}{s_2-1}+\sum_{k=0}^{M-s_2-1} \binom{M}{s_2+k+1}}
\end{align}
The total number of desired symbols is given by
\begin{align}
L(s_2)&=\binom{M-2}{s_2-1}+\sum_{k=1}^{\left\lfloor \frac{M-s_2}{2}\right\rfloor}\binom{M-1}{s_2+2k-1}+\sum_{k=0}^{\left\lfloor \frac{M-s_2-1}{2}\right\rfloor}\binom{M-1}{s_2+2k}\\
&=\binom{M-2}{s_2-1}+\sum_{k=0}^{M-s_2-1}\binom{M-1}{s_2+k}
\end{align}
Thus, the following rate is achievable for $N=2$ and arbitrary $M$
\begin{align}
R(s_2)=\frac{L(s_2)}{t_1(s_2)+t_2(s_2)}=\frac{\binom{M-2}{s_2-1}+\sum_{k=0}^{M-s_2-1}\binom{M-1}{s_2+k}}{M\binom{M-2}{s_2-1}+\sum_{k=0}^{M-s_2-1} \binom{M}{s_2+k+1}}
\end{align}

\section{Further Examples}
In this section, we present further examples to clarify the achievable scheme for some additional tractable values of $M$, $N$.

\subsection{$M=4$ Messages, $N=2$ Databases}
In this example, we show that the achievable rate $R(\tau_2)$ does not match the upper bound $\bar{C}(\tau_2)$ for all traffic ratios $\tau_2$. For $M=4$, we have $M+1=5$ corner points, corresponding to $s_2 =\{0, 1, 2, 3\}$ and another corner point corresponding to the trivial scheme of downloading the contents of database 1. Let $a_i, b_i, c_i, d_i$ denote the randomly permuted symbols of messages $W_1, W_2, W_3, W_4$, respectively. Then, $R(0)=\frac{1}{4}$ by trivially downloading $a_1,b_1,c_1,d_1$ from database 1. In addition, $R(\frac{1}{2})=\frac{1-\frac{1}{2}}{1-(\frac{1}{2})^4}=\frac{8}{15}$ using the symmetric scheme in \cite{JafarPIR}.

\paragraph{Corner Point $s_2=1$:} (See the query table in Table~\ref{M4N2s1}.) The user uses 1 bit of side information in database 2, hence the user starts downloading from round 2 (that admits 2-sums). The user exploits the side information generated in round 1 by downloading $a_2+b_1$, $a_3+c_1$, and $a_4+d_1$. The user completes the stage by downloading undesired symbols consisting of 2-sums that do not include $a_i$, hence the user downloads $b_2+c_2$, $b_3+d_2$, $c_3+d_3$. The undesired symbols are exploited in database 1, thus the user downloads $a_5+b_2+c_2$, $a_6+b_3+d_2$, and $a_7+c_3+d_3$. The user completes the stage by downloading $b_4+c_4+d_4$, which can be exploited in database 2 by downloading $a_8+b_4+c_4+d_4$. In this case, the user downloads $8$ symbols from database 1 and $7$ symbols from database 2, hence we have $\tau_2=\frac{7}{15}$. Since the user downloads $L=8$ desired symbols, the achievable rate $R(\frac{7}{15})=\frac{8}{15}$.

\begin{table}[h]
	\centering
	\caption{The query table for $M=4$, $N=2$, $s_2=1$ (corresponding to $\tau_2=\frac{7}{15}$).}
	\label{M4N2s1}
	\begin{tabular}{|c|c|}
		\hline
		Database 1 & Database 2 \\
		\hline
		$a_1,b_1,c_1,d_1$ & \\
		\hline
		&$a_2+b_1$  \\
		&$a_3+c_1$  \\
		&$a_4+d_1$  \\
		&$b_2+c_2$  \\
		&$b_3+d_2$  \\
		&$c_3+d_3$  \\
		\hline
		$a_5+b_2+c_2$&\\
		$a_6+b_3+d_2$&\\
		$a_7+c_3+d_3$&\\
		$b_4+c_4+d_4$&\\
		\hline
		& $a_8+b_4+c_4+d_4$\\
		\hline
	\end{tabular}
\end{table}

\paragraph{Corner Point $s_2=2$:} (See the query table in Table~\ref{M4N2s2}.) The user downloads $\binom{M-2}{s_2-1}=2$ stages of individual symbols (1-sum) from database 1, so that the user forms 2-sums that can be used in database 2 as side information to start round 3 directly, i.e., by forming 2-sums as side information from the individual symbols, the user effectively skips round 2. More specifically, the user downloads $a_3+b_1+c_1$, $a_4+b_2+d_1$, $a_5+c_2+d_2$ from database 2 taking into considerations that all these undesired symbols are decodable from database 1. The user completes the stage by downloading $b_3+c_3+d_3$ that can be further exploited in database 1 by downloading $a_6+b_3+c_3+d_3$. In this case, the user downloads $9$ symbols from database 1 and $4$ symbols from database 2, therefore $\tau_2=\frac{4}{13}$. The user downloads $L=6$ desired symbols, thus, $R(\frac{4}{13})=\frac{6}{13}$.

\begin{table}[h]
	\centering
	\caption{The query table for $M=4$, $N=2$, $s_2=2$ (corresponding to $\tau_2=\frac{4}{13}$).}
	\label{M4N2s2}
	\begin{tabular}{|c|c|}
		\hline
		Database 1 & Database 2 \\
		\hline
		$a_1,b_1,c_1,d_1$ & \\
		$a_2,b_2,c_2,d_2$ & \\
		\hline
		&\\
		\hline
		&$a_3+b_1+c_1$ \\
		&$a_4+b_2+d_1$\\
		&$a_5+c_2+d_2$\\
		&$b_3+c_3+d_3$\\
		\hline
		 $a_6+b_3+c_3+d_3$&\\
		\hline
	\end{tabular}
\end{table}

\paragraph{Corner Point $s_2=3$:} (See the query table in Table~\ref{M4N2s3}.) In this case, the user skips rounds 2, 3 and jumps directly to round 4 at database 2. Therefore, the user downloads $a_2+b_1+c_1+d_1$ from database 2, which uses $b_1+c_1+d_1$ as side information which is decodable from database 1. Thus, we have $\tau_2=\frac{1}{5}$, and the corresponding rate $R(\frac{1}{5})=\frac{2}{5}$.
\begin{table}[h]
	\centering
	\caption{The query table for $M=4$, $N=2$, $s_2=3$ (corresponding to $\tau_2=\frac{1}{5}$).}
	\label{M4N2s3}
	\begin{tabular}{|c|c|}
		\hline
		Database 1 & Database 2 \\
		\hline
		$a_1,b_1,c_1,d_1$ & \\
		\hline
		&\\
		\hline
			&\\
			\hline
		&$a_2+b_1+c_1+d_1$\\
		\hline
	\end{tabular}
\end{table}

\begin{figure}[t]
	\centering
	\includegraphics[width=0.75\textwidth]{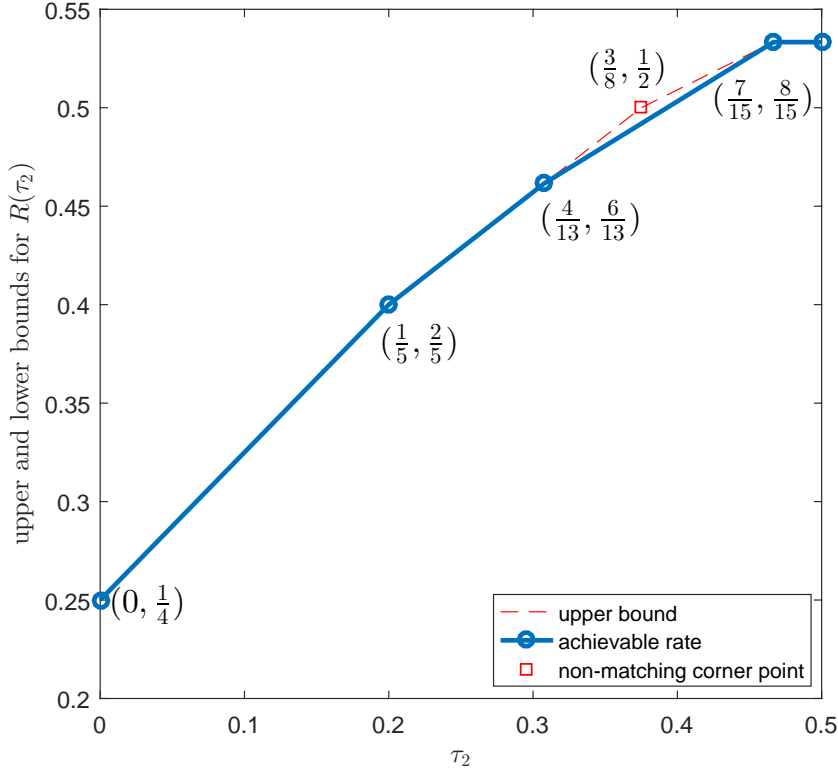}
	\caption{Upper and lower bounds for $R(\tau_2)$ for $M=4$, $N=2$.}
	\label{Fig:M4N2}
	\vspace*{-0.4cm}
\end{figure}

\paragraph{Comparison with the Upper Bound:} The upper bound in Theorem~\ref{Thm1} can be explicitly expressed as:
\begin{align}
R(\tau_2) \leq
\left\{
\begin{array}{ll}
\frac{1}{4}+\frac{3\tau_2}{4}, & 0 \leq \tau_2 \leq \frac{1}{5} \\
\frac{2}{7}+\frac{4\tau_2}{7}, & \frac{1}{5} \leq \tau_2 \leq \frac{3}{8} \\
\frac{4}{11}+\frac{4\tau_2}{11}, & \frac{3}{8} \leq \tau_2 \leq \frac{7}{15} \\
\frac{8}{15}, & \frac{7}{15} \leq \tau_2 \leq \frac{1}{2}
\end{array}
\right.
\end{align}
We observe that for all the corner points of the achievable scheme, the upper and lower bounds match. However, the upper bound has an extra corner point $(\frac{3}{8}, \frac{1}{2})$ which is not achievable using time-sharing. This is illustrated in Fig.~\ref{Fig:M4N2}

\subsection{$M=3$ Messages, $N=3$ Databases}\label{M3N3_scheme}
In this example, we show the capacity-achieving scheme for $M=3$, $N=3$ (the capacity region is illustrated in Fig.~\ref{Fig:M3N3} as a function of $C(\lambda_2,\lambda_3)$). Let $a_i,b_i,c_i$ denote the permuted symbols of messages $W_1,W_2,W_3$, respectively. We show here only the query tables for achieving non-trivial corner points. In this case, we have $\binom{M+N-1}{M}=10$ corner points corresponding to non-decreasing sequences $(n_0,n_1,n_2)$.

For the pair $(\tau_2,\tau_3)=(0,0)$, the achievable scheme is the trivial scheme that downloads $a_1,b_1,c_1$ from the first database only achieving $R(0,0)=\frac{1}{3}$. For the corner point $(\frac{1}{4}, 0)$, this is exactly the same corner point presented in Section~\ref{M3N2 corner} (for $\lambda_2=\frac{1}{3}$) as $\tau_3=0$, which effectively reduces the problem to $N=2$ databases. The achievable scheme for this corner point is illustrated in Table~\ref{c2=1/3}, hence $R(\frac{1}{4}, 0)=\frac{1}{2}$. For the corner point $(\frac{3}{7},0)$, again this point reduces to 2 databases. The achievable scheme is given in Table~\ref{c2=3/4}, and $R(\frac{3}{7}, 0)=\frac{4}{7}$. For the corner point $(\frac{1}{3},\frac{1}{3} )$, which is the symmetric-traffic point, the achievable scheme is the symmetric scheme in \cite{JafarPIR}, which achieves $R(\frac{1}{3},\frac{1}{3})=\frac{9}{13}$. For the corner point $(\frac{1}{2},0)$, we can apply the symmetric achievable scheme for $N=2$ databases only as $\tau_3=0$ in this case, hence $R(\frac{1}{2},0)=\frac{4}{7}$.

Now, we focus on the non-trivial corner points. As mentioned previously, the pair $(s_2,s_3)$ is in bijection with the sequence $(n_0,n_1,n_2)$. Therefore, we enumerate the remaining cases using the pair $(s_2,s_3)$.

\paragraph{Corner Point $(s_2,s_3)=(0,1)$:} In this case, the user does not use the side information generated in database 1 within the initial download of database 2 ($s_2=0$), hence the user downloads new individual symbols from database 2. The user uses 1 bit of side information in database 3 in its round of download (round 2). These side information symbols come from database 1 and database 2. The query table for this case is shown in Table~\ref{M3N3s23}. In this case, we have $(\tau_2,\tau_3)=(\frac{9}{26},\frac{4}{13})$, and the achievable rate is $R(\frac{9}{26},\frac{4}{13})=\frac{9}{13}$.

\begin{table}[h]
	\centering
	\caption{The query table for $M=3$, $N=3$, $(s_2,s_3)=(0,1)$ (i.e., $(\tau_2,\tau_3)=(\frac{9}{26},\frac{4}{13})$).}
	\label{M3N3s23}
	\begin{tabular}{|c|c|c|}
		\hline
		Database 1 & Database 2 & Database 3 \\
		\hline
		$a_1,b_1,c_1$ &$a_2,b_2,c_2$ & \\
		\hline
		$a_3+b_2$ & $a_5+b_1$ & $a_7+b_1$ \\
		$a_4+c_2$ & $a_6+c_1$ & $a_8+c_1$\\
		$b_3+c_3$ & $b_4+c_4$ & $b_5+c_5$\\
						   &  & $a_9+b_2$ \\
		                   &  & $a_{10}+c_2$\\
		                   &  & $b_6+c_6$\\
		\hline
		$a_{11}+b_4+c_4$&$a_{14}+b_3+c_3$ &$a_{17}+b_3+c_3$\\
		$a_{12}+b_5+c_5$&$a_{15}+b_5+c_5$ &$a_{18}+b_4+c_4$\\
		$a_{13}+b_6+c_6$&$a_{16}+b_6+c_6$ &\\
		\hline
	\end{tabular}
\end{table}

\paragraph{Corner Point $(s_2,s_3)=(0,2)$:} The user does not exploit the side information generated from database 1 in the first round of download at database 2. The user uses 2 side information symbols simultaneously in the initial round (round 3) of download at database 3. Note that in round 3 database 3 receives side information from rounds 1 and 2 of databases 1 and 2.  The query table for this case is shown in Table~\ref{M3N3s02}. In this case, we have $(\tau_2,\tau_3)=(\frac{7}{18},\frac{2}{9})$, and the achievable rate is $R(\frac{7}{18},\frac{2}{9})=\frac{2}{3}$.

\begin{table}[h]
	\centering
	\caption{The query table for $M=3$, $N=3$, $(s_2,s_3)=(0,2)$ (i.e., $(\tau_2,\tau_3)=(\frac{7}{18},\frac{2}{9})$).}
	\label{M3N3s02}
	\begin{tabular}{|c|c|c|}
		\hline
		Database 1 & Database 2 & Database 3 \\
		\hline
		$a_1,b_1,c_1$ &$a_2,b_2,c_2$ & \\
		\hline
		$a_3+b_2$ & $a_5+b_1$ & \\
		$a_4+c_2$ & $a_6+c_1$ & \\
		$b_3+c_3$ & $b_4+c_4$ & \\
		\hline
		$a_{7}+b_4+c_4$&$a_{8}+b_3+c_3$ &$a_{9}+b_1+c_1$\\
		                              & &$a_{10}+b_2+c_2$\\
		                              & &$a_{11}+b_3+c_3$\\
		                              & &$a_{12}+b_4+c_4$\\
		\hline
	\end{tabular}
\end{table}

\paragraph{Corner Point $(s_2,s_3)=(1,1)$:} In this case, both databases 2 and 3 exploit the side information generated from database 1 in their initial round of download (round 1). The query table for this case is shown in Table~\ref{M3N3s11}. In this case, we have $(\tau_2,\tau_3)=(\frac{4}{13},\frac{4}{13})$, and the achievable rate is $R(\frac{4}{13},\frac{4}{13})=\frac{9}{13}$.

\begin{table}[h]
	\centering
	\caption{The query table for $M=3$, $N=3$, $(s_2,s_3)=(1,1)$ (i.e., $(\tau_2,\tau_3)=(\frac{4}{13},\frac{4}{13})$).}
	\label{M3N3s11}
	\begin{tabular}{|c|c|c|}
		\hline
		Database 1 & Database 2 & Database 3 \\
		\hline
		$a_1,b_1,c_1$ & & \\
		\hline
		& $a_2+b_1$ & $a_4+b_1$ \\
		& $a_3+c_1$ & $a_5+c_1$\\
		& $b_2+c_2$ & $b_3+c_3$\\
		\hline
	    $a_6+b_2+c_2$ & $a_8+b_3+c_3$ & $a_9+b_2+c_2$ \\
	    $a_7+b_3+c_3$ & & \\
		\hline
	\end{tabular}
\end{table}

\paragraph{Corner Point $(s_2,s_3)=(1,2)$:} In this case, database 2 exploits 1 side information in its initial download (round 2), while database 3 skips to round 3 directly. Database 3 receives side information from the round 1 of database 1 and round 2 of database 2. The query table for this case is shown in Table~\ref{M3N3s12}. In this case, we have $(\tau_2,\tau_3)=(\frac{1}{3},\frac{2}{9})$, and the achievable rate is $R(\frac{1}{3},\frac{2}{9})=\frac{2}{3}$.

\begin{table}[h]
	\centering
	\caption{The query table for $M=3$, $N=3$, $(s_2,s_3)=(1,2)$ (i.e., $(\tau_2,\tau_3)=(\frac{1}{3},\frac{2}{9})$).}
	\label{M3N3s12}
	\begin{tabular}{|c|c|c|}
		\hline
		Database 1 & Database 2 & Database 3 \\
		\hline
		$a_1,b_1,c_1$ & & \\
		\hline
		& $a_2+b_1$ & \\
		& $a_3+c_1$ & \\
		& $b_2+c_2$ & \\
		\hline
		$a_4+b_2+c_2$ & & $a_5+b_1+c_1$ \\
		              & & $a_6+b_2+c_2$\\
		\hline
	\end{tabular}
\end{table}

\paragraph{Corner Point $(s_2,s_3)=(2,2)$:} Both databases 2 and 3 skip round 1 and 2 of downloads and go directly to round 3, in which they exploits 2 side information symbols simultaneously. The query table for this case is shown in Table~\ref{M3N3s22}. In this case, we have $(\tau_2,\tau_3)=(\frac{1}{5},\frac{1}{5})$, and the achievable rate is $R(\frac{1}{5},\frac{1}{5})=\frac{3}{5}$.

\begin{table}[h]
	\centering
	\caption{The query table for $M=3$, $N=3$, $(s_2,s_3)=(2,2)$ (i.e., $(\tau_2,\tau_3)=(\frac{1}{5},\frac{1}{5})$).}
	\label{M3N3s22}
	\begin{tabular}{|c|c|c|}
		\hline
		Database 1 & Database 2 & Database 3 \\
		\hline
		$a_1,b_1,c_1$ & & \\
		\hline
		&  & \\
		\hline
		&$a_2+b_1+c_1$ & $a_3+b_1+c_1$ \\
		\hline
	\end{tabular}
\end{table}

\section{Conclusion}
In this paper, we introduced the PIR problem under asymmetric traffic constraints $\bt$. We investigated the fundamental limits of this problem by developing the novel upper bound $\bar{C}(\bt)= \min_{n_1, \cdots, n_{M-1} \in \{1, \cdots, N\}} \frac{1+\frac{\sum_{n=n_1+1}^{N} \tau_n}{n_1}+\frac{\sum_{n=n_2+1}^{N} \tau_n}{n_1n_2}+\cdots+\frac{\sum_{n=n_{M-1}+1}^{N} \tau_n}{n_0 n_1 \cdots n_{M-1}}}{1+\frac{1}{n_1}+\frac{1}{n_1n_2}+\cdots+\frac{1}{n_0 n_1 \cdots n_{M-1}}}$, for some integer sequence $\{n_i\}_{i=1}^{N} \subset \{1, \cdots, N\}^{M-1}$. The upper bound generalizes the converse proof in \cite{JafarPIR}, which inherently utilizes database symmetry. The upper bound is a piece-wise affine function in $\bt$. The upper bound implies a strict capacity loss due to the asymmetric traffic constraints for certain cases. We developed explicit achievable schemes for $\binom{M+N-1}{M}$ corner points, and achieved the remaining points by time-sharing. We described the achievable scheme by means of a system of difference equations. We explicitly derived the achievable rate for $N=2$ and arbitrary $M$. We proved that the upper bound and the lower bound exactly match for every $\bt$ for the cases of $M=2$ and $M=3$ for any $N$.

\bibliographystyle{unsrt}
\bibliography{references}

\begin{thebibliography}{10}

\bibitem{ChorPIR}
B.~Chor, E.~Kushilevitz, O.~Goldreich, and M.~Sudan.
\newblock Private information retrieval.
\newblock {\em Journal of the ACM}, 45(6):965--981, 1998.

\bibitem{PIRsurvey2004}
W.~Gasarch.
\newblock A survey on private information retrieval.
\newblock In {\em Bulletin of the EATCS}, 2004.

\bibitem{cachin1999computationally}
C.~Cachin, S.~Micali, and M.~Stadler.
\newblock Computationally private information retrieval with polylogarithmic
  communication.
\newblock In {\em International Conference on the Theory and Applications of
  Cryptographic Techniques}. Springer, 1999.

\bibitem{ostrovsky2007survey}
R.~Ostrovsky and W.~Skeith III.
\newblock A survey of single-database private information retrieval: Techniques
  and applications.
\newblock In {\em International Workshop on Public Key Cryptography}, pages
  393--411. Springer, 2007.

\bibitem{yekhanin2010private}
S.~Yekhanin.
\newblock Private information retrieval.
\newblock {\em Communications of the ACM}, 53(4):68--73, 2010.

\bibitem{RamchandranPIR}
N.~B. Shah, K.~V. Rashmi, and K.~Ramchandran.
\newblock One extra bit of download ensures perfectly private information
  retrieval.
\newblock In {\em IEEE ISIT}, June 2014.

\bibitem{unsynchonizedPIR}
G.~Fanti and K.~Ramchandran.
\newblock Efficient private information retrieval over unsynchronized
  databases.
\newblock {\em IEEE Journal of Selected Topics in Signal Processing},
  9(7):1229--1239, October 2015.

\bibitem{YamamotoPIR}
T.~Chan, S.~Ho, and H.~Yamamoto.
\newblock Private information retrieval for coded storage.
\newblock In {\em IEEE ISIT}, June 2015.

\bibitem{VardyConf2015}
A.~Fazeli, A.~Vardy, and E.~Yaakobi.
\newblock Codes for distributed {PIR} with low storage overhead.
\newblock In {\em IEEE ISIT}, June 2015.

\bibitem{RazanPIR}
R.~Tajeddine and S.~El Rouayheb.
\newblock Private information retrieval from {MDS} coded data in distributed
  storage systems.
\newblock In {\em IEEE ISIT}, July 2016.

\bibitem{JafarConf2016}
H.~Sun and S.~A. Jafar.
\newblock The capacity of private information retrieval.
\newblock In {\em IEEE Globecom}, December 2016.

\bibitem{JafarPIR}
H.~Sun and S.~A. Jafar.
\newblock The capacity of private information retrieval.
\newblock {\em IEEE Transactions on Information Theory}, 63(7):4075--4088, July
  2017.

\bibitem{JafarColluding}
H.~Sun and S.~Jafar.
\newblock The capacity of robust private information retrieval with colluding
  databases.
\newblock 2016.
\newblock Available at arXiv:1605.00635.

\bibitem{symmetricPIR}
H.~Sun and S.~Jafar.
\newblock The capacity of symmetric private information retrieval.
\newblock 2016.
\newblock Available at arXiv:1606.08828.

\bibitem{KarimCoded}
K.~Banawan and S.~Ulukus.
\newblock The capacity of private information retrieval from coded databases.
\newblock {\em IEEE Transactions on Information Theory}.
\newblock Submitted September 2016. Also available at arXiv:1609.08138.

\bibitem{arbmsgPIR}
H.~Sun and S.~Jafar.
\newblock Optimal download cost of private information retrieval for arbitrary
  message length.
\newblock 2016.
\newblock Available at arXiv:1610.03048.

\bibitem{codedsymmetric}
Q.~Wang and M.~Skoglund.
\newblock Symmetric private information retrieval for {MDS} coded distributed
  storage.
\newblock 2016.
\newblock Available at arXiv:1610.04530.

\bibitem{MultiroundPIR}
H.~Sun and S.~Jafar.
\newblock Multiround private information retrieval: Capacity and storage
  overhead.
\newblock 2016.
\newblock Available at arXiv:1611.02257.

\bibitem{codedcolluded}
R.~Freij-Hollanti, O.~Gnilke, C.~Hollanti, and D.~Karpuk.
\newblock Private information retrieval from coded databases with colluding
  servers.
\newblock 2016.
\newblock Available at arXiv:1611.02062.

\bibitem{codedcolludedJafar}
H.~Sun and S.~Jafar.
\newblock Private information retrieval from {MDS} coded data with colluding
  servers: Settling a conjecture by {F}reij-{H}ollanti et al.
\newblock 2017.
\newblock Available at arXiv: 1701.07807.

\bibitem{arbitraryCollusion}
R.~Tajeddine, O.~W. Gnilke, D.~Karpuk, R.~Freij-Hollanti, C.~Hollanti, and
  S.~El Rouayheb.
\newblock Private information retrieval schemes for coded data with arbitrary
  collusion patterns.
\newblock 2017.
\newblock Available at arXiv:1701.07636.

\bibitem{MPIRjournal}
K.~Banawan and S.~Ulukus.
\newblock Multi-message private information retrieval: Capacity results and
  near-optimal schemes.
\newblock {\em IEEE Transactions on Information Theory}.
\newblock Submitted February 2017. Also available at arXiv:1702.01739.

\bibitem{codedcolludingZhang}
Y.~Zhang and G.~Ge.
\newblock A general private information retrieval scheme for {MDS} coded
  databases with colluding servers.
\newblock 2017.
\newblock Available at arXiv: 1704.06785.

\bibitem{MPIRcodedcolludingZhang}
Y.~Zhang and G.~Ge.
\newblock Multi-file private information retrieval from {MDS} coded databases
  with colluding servers.
\newblock 2017.
\newblock Available at arXiv: 1705.03186.

\bibitem{BPIRjournal}
K.~Banawan and S.~Ulukus.
\newblock The capacity of private information retrieval from {B}yzantine and
  colluding databases.
\newblock {\em IEEE Transactions on Information Theory}.
\newblock Submitted June 2017. Also available at arXiv:1706.01442.

\bibitem{symmetricByzantine}
Q.~Wang and M.~Skoglund.
\newblock Secure symmetric private information retrieval from colluding
  databases with adversaries.
\newblock 2017.
\newblock Available at arXiv:1707.02152.

\bibitem{tandon2017capacity}
R.~Tandon.
\newblock The capacity of cache aided private information retrieval.
\newblock 2017.
\newblock Available at arXiv: 1706.07035.

\bibitem{wang2017linear}
Q.~Wang and M.~Skoglund.
\newblock Linear symmetric private information retrieval for {MDS} coded
  distributed storage with colluding servers.
\newblock 2017.
\newblock Available at arXiv:1708.05673.

\bibitem{kadhe2017private}
S.~Kadhe, B.~Garcia, A.~Heidarzadeh, S.~El Rouayheb, and A.~Sprintson.
\newblock Private information retrieval with side information.
\newblock 2017.
\newblock Available at arXiv:1709.00112.

\bibitem{wei2017fundamental}
Y.-P. Wei, K.~Banawan, and S.~Ulukus.
\newblock Fundamental limits of cache-aided private information retrieval with
  unknown and uncoded prefetching.
\newblock 2017.
\newblock Available at arXiv:1709.01056.

\bibitem{chen2017capacity}
Z.~Chen, Z.~Wang, and S.~Jafar.
\newblock The capacity of private information retrieval with private side
  information.
\newblock 2017.
\newblock Available at arXiv:1709.03022.

\bibitem{wei2017capacity}
Y.-P. Wei, K.~Banawan, and S.~Ulukus.
\newblock The capacity of private information retrieval with partially known
  private side information.
\newblock 2017.
\newblock Available at arXiv:1710.00809.

\bibitem{sun2017_computation}
H.~Sun and S.~A. Jafar.
\newblock The capacity of private computation.
\newblock 2017.
\newblock Available at arXiv:1710.11098.

\bibitem{mirmohseni2017private}
M.~Mirmohseni and M.~A. Maddah-Ali.
\newblock Private function retrieval.
\newblock 2017.
\newblock Available at arXiv:1711.04677.

\bibitem{abdul2017private}
M.~Abdul-Wahid, F.~Almoualem, D.~Kumar, and R.~Tandon.
\newblock Private information retrieval from storage constrained
  databases--coded caching meets {PIR}.
\newblock 2017.
\newblock Available at arXiv:1711.05244.

\bibitem{wei2017fundamental_partial}
Y.-P. Wei, K.~Banawan, and S.~Ulukus.
\newblock Cache-aided private information retrieval with partially known
  uncoded prefetching: Fundamental limits.
\newblock 2017.
\newblock Available at arXiv:1712.07021.

\end{thebibliography}
\end{document}